\documentclass[superscriptaddress,longbibliography,twocolumn]{revtex4-2}

\usepackage{cancel,soul,ulem}

\usepackage{xcolor}
\usepackage{amsmath,amssymb,graphicx}
\usepackage{graphicx}
\usepackage{epstopdf}
\usepackage{mathtools}
\usepackage{upgreek}
\usepackage{notes2bib}
\usepackage{bm}
\usepackage{color}
\usepackage{braket}
\usepackage{textgreek}
\usepackage{soul}
\usepackage{lipsum}
\usepackage{lineno}
\setstcolor{red}

\begin{document}


\title{Incoherence-assisted mode excitation in non-Hermitian resonant systems}

\author{Amin Hashemi}
\email[Corresponding author: ]{am960769@ucf.edu}
\affiliation{CREOL, The College of Optics and Photonics, University of Central Florida, Orlando, FL 32816, USA}

\author{Vinzenz Zimmermann}
\affiliation{CREOL, The College of Optics and Photonics, University of Central Florida, Orlando, FL 32816, USA}

\author{Armando Perez-Leija}
\affiliation{Department of Electrical and Computer Engineering, Saint Louis University, St. Louis, Missouri 63103, USA}

\author{Andrea Blanco-Redondo}
\email[Corresponding author: ]{andrea.blancoredondo@ucf.edu}
\affiliation{CREOL, The College of Optics and Photonics, University of Central Florida, Orlando, FL 32816, USA}

\begin{abstract}
We introduce and experimentally demonstrate an approach for selective mode excitation in non-Hermitian resonant systems using incoherent light. This method eliminates the need for precise phase control that is often required in coherent excitation schemes. Using this technique on a silicon photonic platform with coupled ring resonators, we successfully excite the topological edge state of a non-Hermitian Su-Schrieffer-Heeger (SSH) model. Our work shows that incoherence-assisted excitation is a robust and passive strategy for topological state preparation, which broadens the scope of non-Hermitian topological photonics thereby providing a practical and experimentally viable tool for selective mode excitation.
\end{abstract}

\maketitle


\section{\label{sec:intro} Introduction}
Non-Hermitian systems, characterized by their ability to exchange energy with the environment, provide a more accurate representation of many real-world physical systems. Non-Hermiticity is not merely an unavoidable feature in certain systems, such as lasers and sensors--where performance inherently depends on energy exchange--but also gives rise to unique phenomena that have no counterparts in Hermitian physics. Examples include exceptional points \cite{Ozdemir2019NatMat, Meng2024APL}, non-orthogonality of eigenmodes \cite{Wonjoo2004IEEE}, loss-induced topological phase transitions \cite{Takata2018PysRevLett,Liu2020PhysRevAppl,Teo2022PhyRevA, Hashemi2025NatMat}, and the non-Hermitian skin effect \cite{Hatano1996PhysRevLett,Yao2018PhysRevLett,Zhang31122022}. These phenomena have been widely explored for their fundamental implications and potential applications in optics \cite{Price2022JPPhotonics,Yan2023Nanophotonics,Nasari2023OptMaterExpress,Hashemi2025APLPhotonics}. 

A central aspect of many non-Hermitian systems is the selective excitation of specific modes, which plays a crucial role in both fundamental studies and practical applications. In non-Hermitian topological models, for instance, exciting edge modes enables robust transport properties and protection against specific types of disorders \cite{Andrea2016PhysRevLett}. In quantum optics, precise mode excitation in waveguide arrays and resonators is essential for efficient generation and manipulation of entangled photon pairs \cite{doi:10.1126/science.aau4296,PRXQuantum.6.010338}, which in non-Hermitian lattices can be critical for quantum sensing \cite{Koch2022PysRevRes,McDonald2020NatComm}. Additionally, in gain-loss engineered structures, selective mode excitation facilitates the exploration of exceptional points and parity-time symmetry breaking \cite{Ozdemir2019NatMat,doi:10.1126/science.aar7709}, leading to unconventional light dynamics.

Among the different platforms for the implementation of non-Hermitian systems, coupled resonator lattices play an important role in optics and photonics \cite{Vahalabook,Vanbook}. These systems can acquire a non-Hermitian nature by engineering the losses or through coupling to external channels. 
While selective excitation of a single mode in Hermitian resonant systems can often be achieved by simply tuning the input field frequency to match the eigenfrequency of a desired mode, the situation is not trivial for non-Hermitian resonant systems due to
the nonzero decay rate of the modes. Specifically, using a single port excitation with a real-frequency field typically results in a steady-state response involving a linear superposition of multiple modes rather than the exclusive excitation of the desired mode. This challenge underscores the need for strategies to efficiently excite a single mode in non-Hermitian resonant systems.
To address this issue, several strategies have been proposed. One approach involves engineering gain and loss to counteract the decay rate of the desired mode, effectively restoring it to a lossless state, or tailoring dissipation to selectively suppress undesired modes while preserving the desired mode with minimal loss \cite{Vinzenz2025PRA}. Another method relies on using complex-frequency excitations to selectively excite a specific mode \cite{Baranov2017Optica,Longhi2018OptLett,Zhong2020PhysRevResearch}.
However, these approaches introduce additional complexities and may not be feasible in all platforms. For instance, in passive photonic platforms, implementing optical gain is often impractical.
Engineering gain and loss to selectively
counteract the decay of a desired mode requires integrating the system on an active platform, which presents significant practical challenges depending on the intended application.
Alternatively, a complex input frequency that matches the eigenfrequency of the target mode must have a negative imaginary component, implying exponential temporal decay. Consequently, the input power decreases over time. While modulation schemes with carefully shaped amplitude profiles could potentially overcome this limitation, they require precise knowledge of the system and the ability to perform ultrafast amplitude modulation, which may not be feasible in many platforms.
A more feasible alternative is to use a multiport input signal, wherein multiple resonators are simultaneously excited with a real-frequency field. This method enhances control over mode selection and energy distribution while eliminating the need for gain but requires precise control over the relative phases and amplitudes of the multiport input signal.\\

In this work, we propose and experimentally demonstrate an alternative approach based on incoherent excitation of non-Hermitian systems, where the relative phases of a multiport input signal vary randomly in time such that the resulting input is effectively incoherent. Our method offers two significant advantages. First, it enables efficient mode excitation without the need for the precise phase adjustments typically required by coherent excitation schemes. While incoherent excitation does not outperform optimally phase-matched coherent driving, it significantly reduces sensitivity to phase noise and simplifies the experimental setup by removing the need for complex phase-locking hardware.
Second, our approach is remarkably robust: it remains effective even when the excitation signal has suboptimal spatial overlap with the desired mode. This feature further streamlines implementation, making the technique highly practical for a wide range of applications. 

\section{Results}
We consider a non-Hermitian resonant open system in which non-Hermiticity arises due to coupling with external channels, i.e., input/output ports, and the presence of intrinsic loss. Within the framework of linear temporal coupled mode theory (TCMT) \cite{Fan2003josaA}, the light dynamics in this system is governed by the equation
\begin{equation}
\label{eq:tcmt}
i\frac{d\ket{a(t)}}{dt} = \hat{H}\ket{a(t)}+i\hat{\Gamma}\ket{b(t)},
\end{equation}
where $\ket{b(t)}$ represents the input field vector and $\ket{a(t)}$ describes the evolved system state. The matrices $\hat{H}$ and $\hat{\Gamma}$ correspond to the system Hamiltonian and the coupling between the input channels and resonators, respectively.
A general incoherent input, cannot be represented by a single deterministic vector; rather, it must be described as a random function or stochastic process \cite{stochasticprocessesbook}. Since the input field is a randomly varying vector, the corresponding system state $\ket{a(t)}$ also becomes a random process. In this case, Eq.~(\ref{eq:tcmt}) takes the form of a stochastic differential equation (SDE), and determining the system dynamics $\ket{a(t)}$ would, in principle, require methods from stochastic calculus, such as It\^{o} integration \cite{SDEbook}.
However, in this work we focus on a special case where, despite the randomness of the input field, the system dynamics can be analyzed without resorting to SDE techniques. To explain this, we note that resonant systems are characterized by a time constant, $\tau_s$—the characteristic time it takes for the system to reach its steady-state response after excitation by a coherent input (see Appendix \ref{appendix_transient}). Following such an excitation, the system first undergoes a transient evolution, and after approximately one time constant, it settles into a steady state in which the intracavity power no longer changes in time (see Appendix \ref{appendix_transient}).
Now, consider a \textit{spatially incoherent} input field vector, in which the phases of the individual field components evolve randomly in time. The input field is characterized by a finite coherence time $\tau_c$, defined as the timescale over which these phases remain approximately constant. We assume the coherence time, $\tau_c$, is longer than the system’s time constant: $\tau_c>\tau_s$. During the transient evolution toward the steady state, the phases of the individual field components can be regarded as approximately constant, so the system responds as if driven by a coherent input. Once the system reaches steady state, the phases of the input field vector may change randomly, causing the system to enter a new transient period and subsequently reach another steady state corresponding to the new input phases.
We emphasize that these steady states are forced responses of a driven–dissipative system and are not eigenstates of the evolution operator; their phase dependence arises solely from the external drive.
If the detector’s response time, $\tau_d$, is longer than the coherence time of the input, i.e., $\tau_d>\tau_c$, the measured output signal effectively represents an average over multiple steady-state responses corresponding to different random input phases. Although these steady states are random, the temporal dynamics connecting them follows the same deterministic evolution as that of a coherently driven system. Consequently, the randomness observed in the measurement originates from averaging over these distinct steady-state responses during the detector’s integration time.
The conditions discussed above for the coherence time, time constant of the system, and detector response time are readily achievable in practical experiments. For example, in the integrated photonic platform used in our work, the relaxation time of the system is approximately $\tau_s \approx 1$~ns (see Appendix \ref{appendix_transient}). To satisfy the condition $\tau_c > \tau_s$, the incoherent input must have a coherence time longer than 1~ns, corresponding to a bandwidth of about $\Delta\omega \leq 100$~MHz. 
The detector’s integration time should also satisfy $\tau_d > \tau_c$, which can be met using detectors with millisecond-scale response times—readily available in standard photodetectors and power meters.
Therefore, the assumed condition $\tau_s < \tau_c < \tau_d$ is experimentally feasible and provides a means to bridge the deterministic transient dynamics—governed by coherent excitation—with the statistical ensemble of steady states produced under incoherent driving.\\

To investigate the steady-state response of the system, we employ the frequency-domain resolvent, also known as the Green’s operator
\begin{equation}
\label{eq:freq_response}
\ket{A(\omega)}=\hat{G}(\omega)i\hat{\Gamma}\ket{B(\omega)},
\end{equation}
where $\hat{G}(\omega)$ is the Green’s operator $\hat{G}(\omega)\equiv\left(\omega \hat{I}-\hat{H}\right)^{-1}$. In this formulation, $\ket{A(\omega)}=\mathcal{FT}\left(\ket{a(t)}\right)$ and $\ket{B(\omega)}=\mathcal{FT}\left(\ket{b(t)}\right)$ denote the Fourier transforms of the evolved state and the input field, respectively, with $\mathcal{FT}(\cdot)$ representing the Fourier transform operation.
Notice, Eq.~(\ref{eq:freq_response}) applies to coherent input fields with separable time dependence, $\ket{b(t)} = b(t)\ket{u}$. Although incoherent input signals do not strictly meet this condition, the equation remains valid if we consider the incoherent signal as an ensemble average of many coherent signals with random phases \cite{RevModPhys.37.231, statisticalopticsbook}. Therefore, the state of the system under incoherent excitation is a mixed state which can be described either as an ensemble of pure states $\ket{A(\omega)}$ or, equivalently, by a frequency-domain density matrix—also referred to as the coherence matrix—defined as $\tilde{\rho}(\omega)\equiv \braket{\ket{A(\omega)}\bra{A(\omega)}}$, where $\braket{\cdot}$ denotes an ensemble average over realizations of the input field. Using Eq.~(\ref{eq:freq_response}), one can directly show that the system’s coherence matrix is related to the input coherence matrix, $\tilde{\rho}_s$, via the resolvent operator according to $\tilde{\rho}(\omega)=\hat{G}(\omega)\hat{\Gamma}\tilde{\rho}_s(\omega)\hat{\Gamma^\dagger} \hat{G}^{\dagger}(\omega)$, where $\tilde{\rho}_s(\omega)\equiv\braket{\ket{B(\omega)}\bra{B(\omega)}}$ characterizes the statistical properties of the incoherent input field vector.

For non-Hermitian resonant systems that do not exhibit exceptional points --such as the one considered in this study-- the Green's operator can be expressed as \cite{Hashemi2022NatComm}
\begin{equation}
\label{eq:green_operator}
\hat{G}(\omega)=\sum_n\frac{\ket{\psi_n^r}\bra{\psi_n^l}}{\omega-\Omega_n},
\end{equation}
where $\ket{\psi_n^r}$ ($\bra{\psi_n^l}$) are the right (left) eigenvectors of the Hamiltonian satisfying 
$\hat{H}\ket{\psi_n^r}=\Omega_n\ket{\psi_n^r}$ ($\bra{\psi_n^l}\Omega_{n}^{*}=\bra{\psi_n^l}\hat{H}^{\dagger}$).
The left and right eigenvectors are normalized through the biorthogonality relation, $\braket{\psi_n^l|\psi_m^r}=\delta_{nm}$.
The corresponding eigenvalues, $\Omega_n$, represent the system's eigenfrequencies, with their imaginary parts indicating the decay (Im$(\Omega_n)<0$) or growth (Im$(\Omega_n)>0$) rate of the associated modes.

We note that the system response may also be described within a time-domain framework, as governed by the following master equation:
\begin{align}\label{eq:master}
    \frac{d\hat{\rho}(t)}{dt}=-i\left(\hat{H}\hat{\rho}-\hat{\rho}\hat{H}^\dagger\right) + \hat{D}(t),
\end{align}
where $\hat{\rho}(t)\equiv\braket{\ket{a(t)}\bra{a(t)}}$ denotes the time-domain coherence matrix of the system state, and 
\begin{align}\label{eq:master_source}
    \hat{D}(t)\equiv\int_0^t\hat{U}(t-t')\hat{\Gamma} \hat{\rho}_s(t',t)\hat{\Gamma}^\dagger dt' + h.c.
\end{align}
represents the input signal or the drive contribution. Here $\hat{\rho}_s(t',t)\equiv\braket{\ket{b(t')}\bra{b(t)}}$ denotes the time-domain coherence matrix of the input field and $\hat{U}(t)\equiv e^{-i\hat{H}t}$ is the system evolution operator. A detailed derivation of Eq.~(\ref{eq:master}) is provided in Appendix \ref{appendix_master_eq}.
To determine the steady-state response of the system within the time-domain framework, we impose the stationarity condition $\frac{d\hat{\rho}}{dt}=0$ and consider the asymptotic limit $t\rightarrow \infty$. Under these conditions, Eq.~(\ref{eq:master}) reduces to the Lyapunov equation:
\begin{align}
    \left(-i\hat{H}\right)\hat{\rho}+\hat{\rho}\left(-i\hat{H}\right)^\dagger + \hat{D}(\infty)=0
\end{align}
 whose solution uniquely determines the steady-state response of the system.
In the case of a narrow-band incoherent input with a well-defined central frequency $\omega$, as shown in detail in Appendix \ref{appendix_master_eq}, the coherence matrix is given by:
\begin{align}\label{eq:lyanunov_source}
    \hat{D}(\infty)=i\hat{G}(\omega)\hat{\Gamma}\hat{S}(\omega)\hat{\Gamma}^\dagger + h.c.
\end{align}
where $\hat{S}(\omega)$ denotes the power spectral density of the input field. Equation~(\ref{eq:lyanunov_source}) can be applied to a general input signal with an arbitrary spectral profile to determine the asymptotic drive term $\hat{D}(\infty)$. For example, as shown in Appendix \ref{appendix_master_eq}, in the case of a fully incoherent input signal with a broadband spectrum uniformly covering all frequencies, the drive term reduces to $\hat{D}=\hat{\Gamma}\hat{S}\hat{\Gamma}^\dagger$.
 In the remainder of this paper, we employ a frequency-domain framework to analyze the system, as it provides enhanced physical insight into the parameters that control the selective excitation of specific system modes, along with their spectral and resonance characteristics.

Based on Eqs.~(\ref{eq:freq_response}) and (\ref{eq:green_operator}), the excitation of a given mode—under either coherent or incoherent
excitation—is governed by three primary factors: (i)  the decay rate of the mode, (ii) the spatial overlap between the input signal and the target mode, and (iii) the real part of the mode’s eigenfrequency relative to the input frequency. In general, an input signal with frequency $\omega$ can excite all the eigenmodes $\ket{A(\omega)}=\sum_n c_n(\omega) \ket{\psi_n^r}$, with amplitude coefficients $c_n(\omega)=\frac{\braket{\psi_n^l|i\hat{\Gamma}|B(\omega)}}{\omega-\Omega_n}$.
Accordingly, if a particular mode, say mode \#$m$, exhibits no decay, $\text{Im}(\Omega_m)=0$, then tuning the input frequency to the mode’s eigenfrequency, $\omega=\Omega_m$, and ensuring a nonzero overlap between the input signal and the mode’s left eigenvector, $\braket{\psi_m^l|i\hat{\Gamma}|B(\Omega_m)}\neq 0$, will predominantly excite mode \#$m$, while keeping the excitation of other modes negligible, as $c_m(\Omega_m)\rightarrow\infty$. However, in reality, $c_m(\Omega_m)$ does not diverge indefinitely; instead, nonlinear effects emerge, governing the system dynamics and preventing unbounded growth.
This scenario mirrors the selective excitation of a single mode in a Hermitian resonant system, where tuning the input frequency to match the eigenfrequency of a mode with zero decay rate ensures that the target mode dominates the system response while contributions from other modes remain negligible. We refer to this as the \textit{efficient excitation} of a single desired mode.
However, for passive systems (i.e., systems without any optical gain) with intrinsic loss, all system modes exhibit a finite decay rate, i.e.,
$\text{Im}[\Omega_n]\neq 0$. Consequently, tuning the input frequency to match the real part of a mode’s eigenfrequency, along with ensuring a nonzero overlap between the input and the mode’s left eigenvector, is not sufficient for efficient excitation of a desired mode. 
In this case, the amplitude coefficient becomes $c_m = \frac{\braket{\psi_m^l|i\hat{\Gamma}|B(\omega)}}{-\text{Im}[\Omega_m]}$, which remains finite and should be contrasted with the Hermitian case where $c_m \rightarrow \infty$ as $\text{Im}(\Omega_m) \rightarrow 0$.
Therefore, the denominator of the mode amplitude coefficients $c_m$ in non-Hermitian passive systems is nonzero and, depending on the decay rate of each mode, can deviate significantly from zero. As a result, efficient excitation cannot be achieved solely through frequency tuning. To enhance the value of $c_m$ and thereby improve the excitation efficiency of a specific mode (say, mode \#$m$), one must increase the numerator of $c_m$, i.e., the overlap term $\braket{\psi_m^l|i\hat{\Gamma}|B(\omega)}$. This requires engineering the spatial profile of the input field to better match the field coupled to the resonators, i.e., $i\hat{\Gamma}\ket{B(\omega)}$, with the eigenvector of the desired mode.
Using multiple input ports can enhance the overlap of the input signal with the desired mode, defined as $\braket{\psi_m^l|i\hat{\Gamma}|B(\omega)}$, thereby improving excitation efficiency. However, precise phase adjustment of the input signals is crucial, as improper phase tuning can further degrade excitation efficiency.
Additionally, selective excitation becomes challenging when the real parts of the eigenfrequencies of two modes are very close to each other—specifically, when their separation is smaller than the linewidth of the input signal. In such a case, both coherent and incoherent excitation schemes are fundamentally limited in their ability to address the modes independently.
\begin{center}
\begin{figure*}
\includegraphics[width=\textwidth]{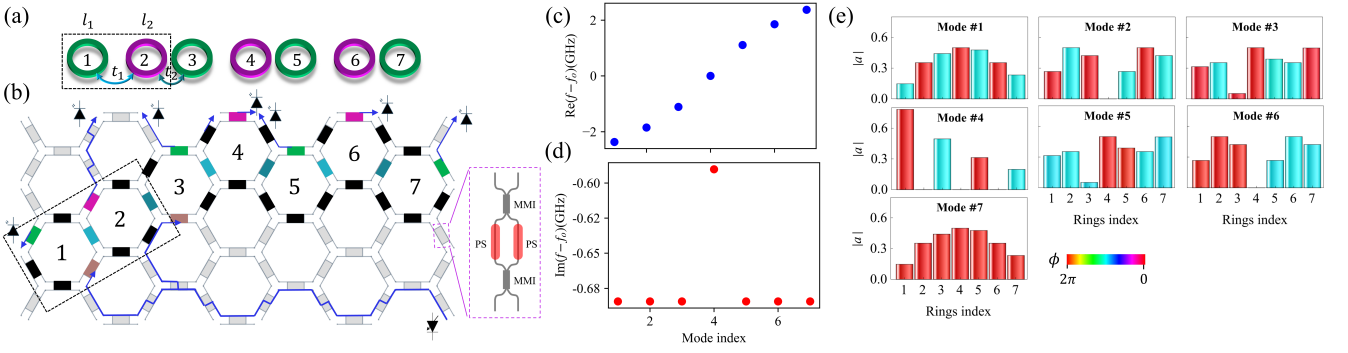}
\caption{(a) Non-Hermitian SSH model implemented using a 1D array of seven ring resonators. Each unit cell (outlined by a dotted rectangle) consists of two rings with unbalanced losses $l_1$ and $l_2$, intracell coupling $t_1$, and intercell coupling $t_2$. (b) Experimental implementation of the ring resonator chain on a programmable integrated photonic platform. Each programmable unit is a Mach-Zehnder interferometer (MZI), as shown in the inset, which consists of two multimode interferometers (MMIs) and two phase shifters (PSs). The loss in each ring is controlled by its coupling to the output port (indicated in green and pink). The coupling between adjacent rings is adjusted via the common MZI shared between them (depicted in dark and light cyan). The input ports for rings \#1 and \#3 are implemented using MZIs (shown in brown), and rings \#3 and \#4 can be excited in a similar manner. (c), (d) Plots of the real and imaginary parts of the eigenfrequencies for the non-Hermitian SSH model implemented using a 1D array of ring resonators. (e) Visualization of the eigenvectors for the non-Hermitian SSH model under consideration, with the color of each bar representing the phase of the mode amplitude at the corresponding ring resonator.}
\label{fig:mesh}
\end{figure*}
\end{center}
To demonstrate this concept, we consider the topological non-Hermitian Su–Schrieffer–Heeger (SSH) model \cite{SSH1979PhysRevLett,Lieu2018PhysRevB}. This model is experimentally realized using a chain of seven microring resonators, shown in Fig.\ref{fig:mesh}(a), on a programmable integrated photonic platform, which consists of a hexagonal mesh of Mach-Zehnder interferometers (MZIs), depicted in Fig.\ref{fig:mesh}(b). Each MZI comprises two multimode interferometers (MMIs) and two phase shifters, as shown in the inset of Fig.\ref{fig:mesh}(b), allowing precise control over both the power splitting ratio and the phase of light at the MZI outputs.
It is worth noting that the incoherent excitation scheme proposed here does not rely on the reconfigurability of the platform; it can, in principle, be implemented on any application-specific (non-reconfigurable) photonic architecture as well.
This configuration enables dynamic routing of light across the chip from the input to the output ports or confinement within a ring resonator, which is formed by six interconnected MZIs. The coupling between adjacent ring resonators is engineered such that the intracell coupling strength is set to $t_1=0.99$ GHz, while the intercell coupling strength is adjusted to $t_2=1.56$ GHz via the shared MZI between two rings.
Each ring resonator is characterized by a quality factor of $Q\approx 4\times 10^4$, an intrinsic loss of $l_o=0.56$ GHz, and a resonance frequency of $f_o=193.5$ THz. To engineer the desired non-Hermitian effects, a staggered loss pattern of $L_1=0.03$ GHz and $L_2=0.22$ GHz is imposed by selectively extracting a fraction of light from alternating rings to the output ports. The total loss of each ring is therefore $l_i=l_0+L_j$, where $i$ is the ring index and $j = {1,2}$. All input ports are coupled to their respective ring resonators with a coupling strength of $k=0.44$ GHz. The experimental characterization of these model parameters is achieved through single-ring and double-ring add-drop filter measurements, as detailed in the appendix of Ref. \cite{Hashemi2025NatMat}.

Figure \ref{fig:mesh}(c), (d) presents the real and imaginary parts of the model's eigenfrequencies, showing that all modes exhibit a nonzero decay rate. Our focus is on the excitation of the edge mode (i.e., mode \#4), which has lower decay rate compared to the other modes and it is topologically protected. If the desired mode had a higher decay rate instead, the concept of incoherent-assisted excitation would still hold, though its impact would be less pronounced—particularly in low-$Q$ ring resonators, such as those used in our experiments. A detailed analysis of a bulk mode excitation, supporting the discussion here, is presented in Appendix~\ref{appendix_bulk_excitation}.

By comparing the power distribution across the ring resonator chain with that of the desired mode (in this case, the edge mode), we can estimate how efficiently the edge mode has been excited. The spatial distribution of the power is represented as  $\mathcal{P}=[p_1, p_2, \cdots, p_7]$, where $p_i$ is power at $i$th ring resonator. To quantify how close the state of the system is to the edge state we define the following parameter, referred to as mismatch index:
\begin{equation}
\label{eq:eta}
\eta = \left|\frac{\mathcal{P}}{|\mathcal{P}|} - \frac{\mathcal{P}_e}{|\mathcal{P}_e|} \right|,
\end{equation}
where $\mathcal{P}_e$ is the power distribution of the edge mode and $\,|\cdot|$ stands for vector norm, i.e., $|\mathcal{P}|=\sqrt{p_1^2+p_2^2+\cdots+p_7^2}$. If the field coupled to the system matches the eigenvector associated with the edge state, the mismatch index is zero. Conversely, as the field coupled to the system deviates further from this eigenvector, the system response increasingly differs from the edge state, leading to a higher mismatch index.

The spatial distribution of field amplitudes across the chain of ring resonators for all the modes are shown in Fig.\ref{fig:mesh} (e). As it can be seen in Fig.\ref{fig:mesh} (e), the edge state (mode \#4) is predominantly localized at ring \#1, with its amplitude vanishing at even-numbered rings. The color of the bars represents the phase of the field at each ring resonator.

\begin{center}
\begin{figure}
\includegraphics[width=2.3in]{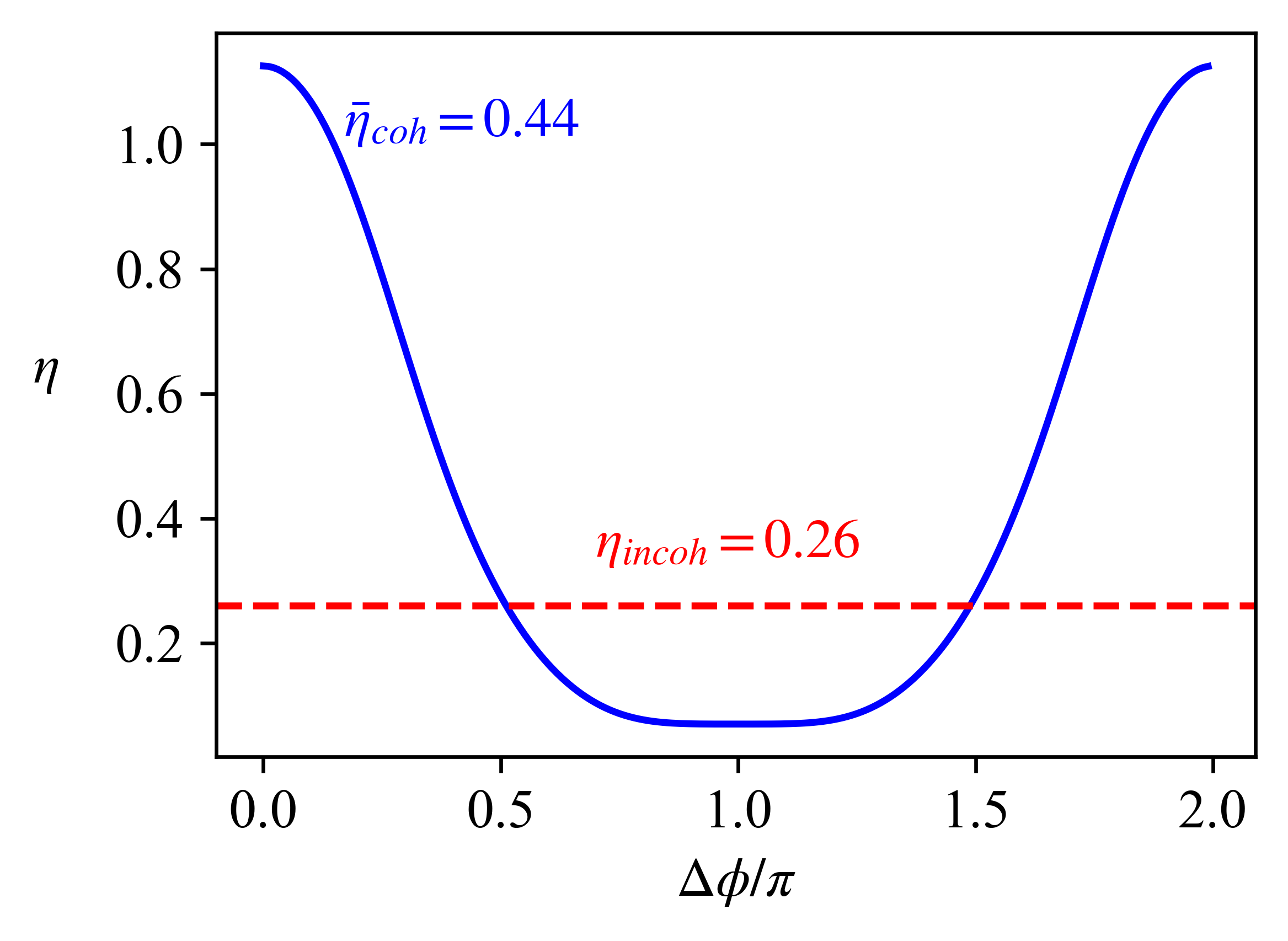}
\caption{Mismatch index $\eta$ for exciting rings \#1 and \#3 with a coherent input signal having a relative phase of $\Delta\phi$, as predicted using TCMT for the experimental model parameters given in the paper. The average mismatch index for coherent excitation is $\bar{\eta}_{coh}=0.44$, while for an incoherent input signal, it is $\eta_{incoh}=0.26$ in this case.}
\label{fig:eta_13}
\end{figure}
\end{center}
\begin{center}
\begin{figure*}
\includegraphics[width=0.9\textwidth]{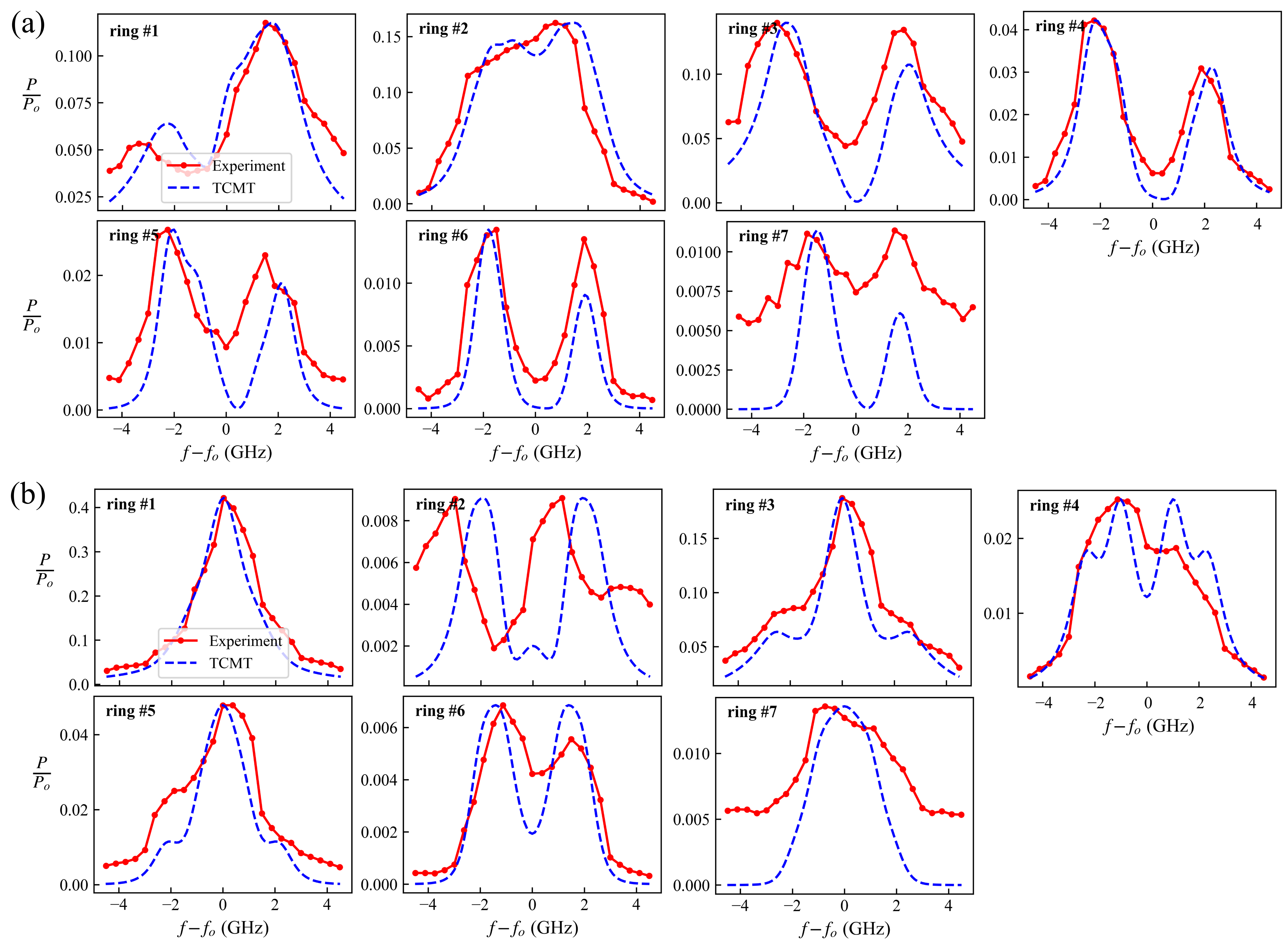}
\caption{Power spectrum of the ring resonators obtained from experimental measurements (red line) and TCMT analysis (dashed blue line) by exciting rings \#1 and \#3 with a coherent input signal having a relative phase of (a) $\Delta\phi=0.15\pi$ and (b) $\Delta\phi=\pi$. The experimental data is normalized so that the maximum power in each spectrum matches the maximum power obtained from TCMT.}
\label{fig:coh_13_spec}
\end{figure*}
\end{center}
\begin{center}
\begin{figure*}
\includegraphics[width=0.9\textwidth]{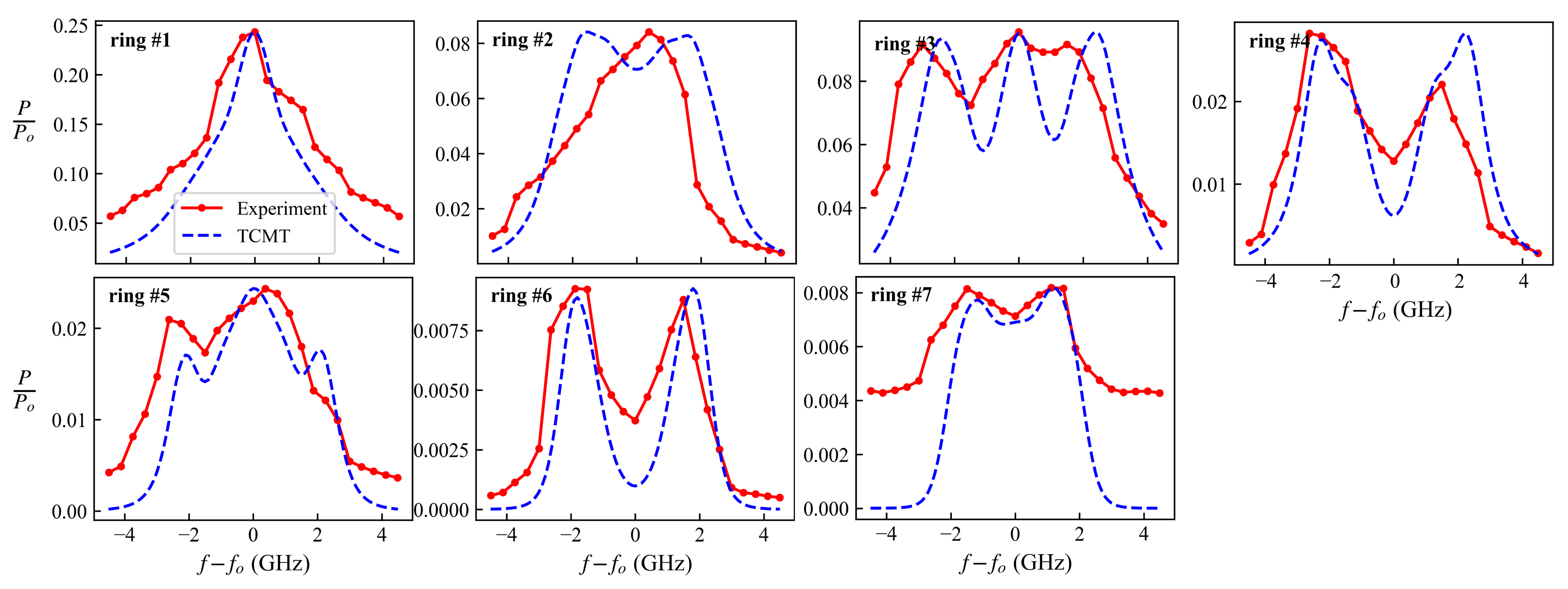}
\caption{Power spectrum of the ring resonators obtained from experimental measurements (red line) and TCMT analysis (dashed blue line) under incoherent excitation of rings \#1 and \#3. Incoherent excitation is achieved by averaging 50 coherent excitations with random relative phases. The experimental data is normalized so that the maximum power in each spectrum matches the maximum power predicted by TCMT.}
\label{fig:incoh_13_spec}
\end{figure*}
\end{center}
\begin{center}
\begin{figure}
\includegraphics[width=3.5in]{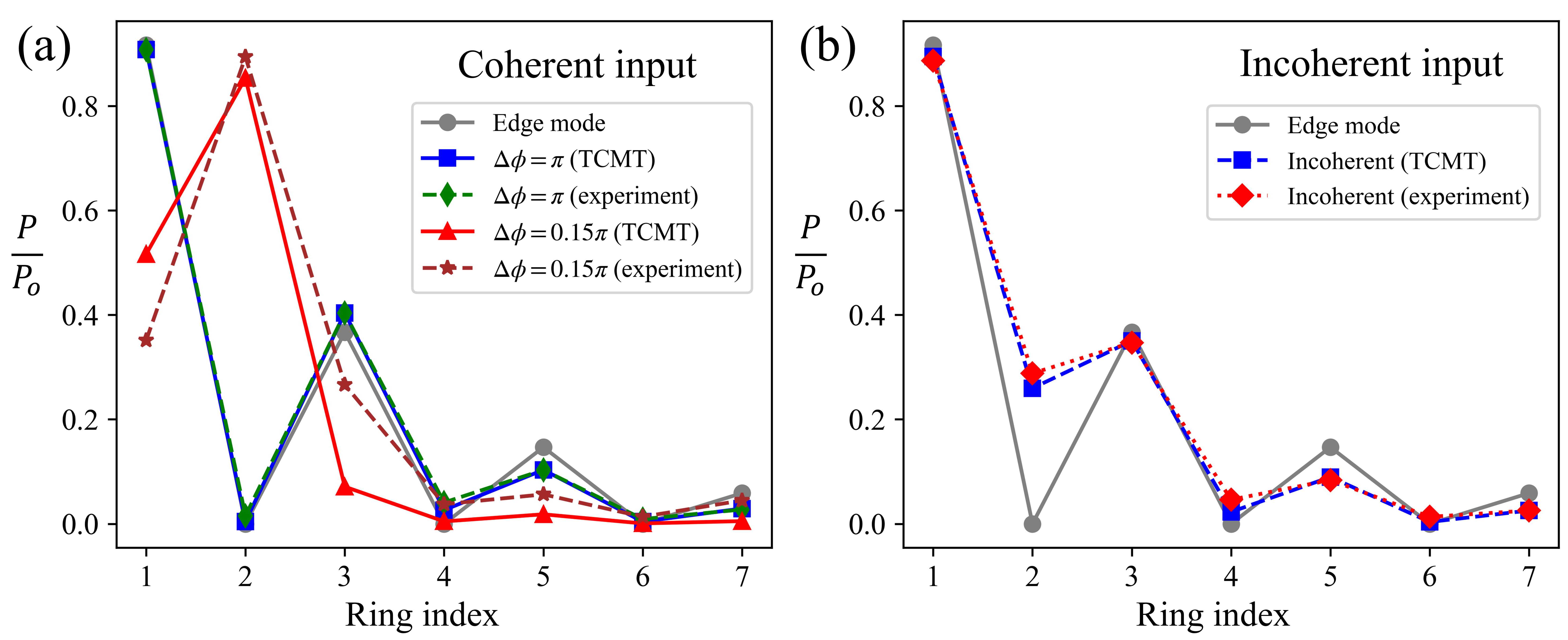}
\caption{(a) Spatial power distribution of the edge state (gray line) compared to the system's spatial power distribution following excitation with a coherent input at frequency $f=f_o$, coupled to rings \#1 and \#3. Experimental measurements and TCMT predictions are shown for two relative input phases, $\Delta\phi=0.15\pi$ and $\Delta\phi=\pi$. (b) Spatial power distribution under incoherent excitation, with experimental results (dotted red line) and TCMT predictions (dashed blue line).}
\label{fig:13_power}
\end{figure}
\end{center}
Let us now consider a situation with strong spatial overlap of the input signal with the desired eigenmode, specifically a two-port input signal, where one port excites ring \#1 and the other excites ring \#3 with equal amplitudes but a relative phase of $\Delta\phi$. To analyze how the system's state evolves as a function of $\Delta\phi$, we plot the mismatch index as a function of the relative input phase when exciting rings \#1 and \#3, shown as the blue line in Fig.\ref{fig:eta_13}. The details of numerical simulation and caluclations are given in Appendix \ref{appendix_numerical_simulations}. As expected, Fig.\ref{fig:eta_13} confirms that the mismatch index reaches its minimum at $\Delta\phi=\pi$ -- the exact phase difference between ring \#1 and \#3 of desired eigenmode (mode \#4) -- and increases progressively as $\Delta\phi$ deviates further from $\pi$.
While exciting ring \#1 and ring \#3 through their respective input ports with a relative phase of $\Delta\phi=\pi$ can achieve strong overlap with the edge state, an incorrect relative phase between the inputs can significantly hinder the efficient excitation of the edge mode.
To further examine this experimentally, we first excite the system with a coherent input, once with a relative phase of $\Delta\phi=0.15\pi$ and then again with a relative phase of $\Delta\phi = \pi$. One of the MZIs on the programmable platform, placed in the path of the input port exciting ring \#3, is used to adjust the relative input phase. To ensure precise phase tuning, we compare the experimental power spectrum of the rings (red solid line) with the TCMT-derived spectrum (dashed blue curve) for each case, as shown in Fig.\ref{fig:coh_13_spec} (a), (b), respectively. This comparison is crucial, as the power spectrum is highly sensitive to the relative phase between the two input ports. The experimental data is normalized so that the maximum power spectrum of each ring aligns with the TCMT-derived maximum, compensating for variations in coupling loss at the input/output ports. After setting the relative phase, we measure the normalized power at each ring resonator for an input signal with a frequency matching the real part of the eigenfrequency of the edge state (mode \#4), i.e., $f = \text{Re}[\Omega_4]/2\pi=f_o$, where $\Omega_4$ represents the eigenfrequency of mode \#4. Figure \ref{fig:13_power} (a) shows the spatial power distribution for a coherent input signal exciting rings \#1 and \#3 with $\Delta\phi=0.15\pi$, obtained from experimental measurements (brown dashed line) and TCMT analysis (red line). The spatial power distributions presented throughout this work are normalized by their vector norm, as defined below Eq.~(\ref{eq:eta}). Additionally, the spatial power distribution for $\Delta\phi = \pi$ is presented, with experimental results (dashed green line) and TCMT predictions (blue line). Fig.\ref{fig:13_power}(a) clearly illustrates how the system state is influenced by varying the relative input phase. An improper choice of relative phase can lead to the excitation of multiple modes instead of exclusively targeting the desired mode. This effect is further reflected in the increased mismatch index for coherent excitation with $\Delta\phi=0.15\pi$, measured experimentally as $\eta_{exp}^{coh}=1.07$, while TCMT analysis predicts $\eta_{TCMT}^{coh}=1.00$.

In contrast, an incoherent input signal, where rings \#1 and \#3 are excited with random relative phases, results in a consistently efficient excitation of the edge state. To generate an incoherent input, we randomly vary the relative phase between the two ports and compute the ensemble average of the measured power distributions. 
It is worth mentioning that equivalent incoherent inputs can be generated externally, for example, by using a fiber-based phase shifter \cite{hashemi2026programmableonchipsynthesisreconstruction} or by introducing enough path-length imbalance \cite{hashemi2026onchipcontrolcoherencematrix} before coupling light into the chip.
For input signals with coherence time $\tau_c$ longer than the system relaxation time $\tau_s$, each realization can be considered quasi-coherent during the transient, allowing the system to reach a steady-state response corresponding to that realization. The detector then effectively averages over many such realizations, reproducing the statistical effect of incoherent driving. This approach is equivalent to using a physical incoherent input with appropriate coherence time.
The average power spectrum of the rings, measured experimentally for 50 input signals with randomly varying relative phases, is shown in Fig. \ref{fig:incoh_13_spec} (red solid line) and compared with the TCMT prediction (dashed blue line). 
The sequence of random-phase pulses doesn't need to occur within a specific time interval as long as $\tau_s<\tau_c<\tau_d$ is satisfied. In this case each random-phase realization of the input can be viewed as a quasi-static excitation that persists long enough for the system to reach a steady state before the phase changes.
From Fig. \ref{fig:incoh_13_spec}, the spatial power distribution for an incoherent input signal at frequency $f=f_o$ is extracted and presented in Fig. \ref{fig:13_power}(b) for both experimental measurements (red dotted line) and TCMT predictions (blue dashed line). The corresponding mismatch indices are $\eta_{exp}^{incoh}=0.30$ for the experimental data and $\eta_{TCMT}^{incoh}=0.26$ from TCMT analysis.
The results indicate that, for the given model parameters, incoherent excitation yields a lower mismatch index compared to the average mismatch index obtained under coherent excitation, $\bar{\eta}_{coh}=0.44$.
This is a meaningful improvemnet of $60\%$ over the phase-averaged coherent case, while remaining within a practically achievable configuration. It should be emphasized that the incoherent excitation scheme provides this improvement without requiring any phase optimization or calibration.
This implies that an incoherent input signal can excite the edge state more efficiently than a coherent input signal with a randomly chosen or imprecisely tuned relative phase.
In realistic experimental settings, however, the degree of phase control is highly system dependent. Variations in fiber length, on-chip routing, thermal fluctuations, and coupling-induced phase drifts can render precise phase stabilization between input ports challenging or impractical \cite{Jorge2024Advanced_photonics_nexus}. A discussion of phase instability in multi-port excitation without active phase control, along with experimental evidence, is provided in Appendix~\ref{appendix_phase_flactuation}. Consequently, we evaluate the average of $\eta$ over random input phases as a phase-independent performance metric, providing an experiment-agnostic benchmark for comparing coherent and incoherent excitation schemes. While specific implementations may define alternative figures of merit based on the achievable level of phase control, the averaged $\bar{\eta}_{coh}$ adopted here offers a robust and general measure of excitation efficiency under realistic conditions.
Moreover, this improvement could be even more pronounced if the intrinsic losses of the ring resonators were further reduced.

As shown in Fig.\ref{fig:eta_13}, a coherent input signal exciting rings \#1 and \#3 with a relative phase of $|\Delta\phi-\pi|<\pi/4$ still ensures efficient edge mode excitation due to the relatively flat behavior of the mismatch index curve around $|\Delta\phi-\pi|<\pi/4$. This occurs because, for all modes except the edge state (mode \#4), the field amplitudes at rings \#1 and \#3 are in phase, whereas only for the edge mode, they are out of phase (see Fig.\ref{fig:mesh} (e)). Consequently, an input signal with $|\Delta\phi-\pi|<\pi/4$ has a strong overlap with the edge state while maintaining minimal overlap with other modes, thereby suppressing their excitation. However, even in this scenario, incoherent excitation proves to be more effective than coherent excitation with a randomly chosen phase. The mismatch index for incoherent excitation remains significantly lower, highlighting its superior mode selectivity. 
\begin{center}
\begin{figure}
\includegraphics[width=3.5in]{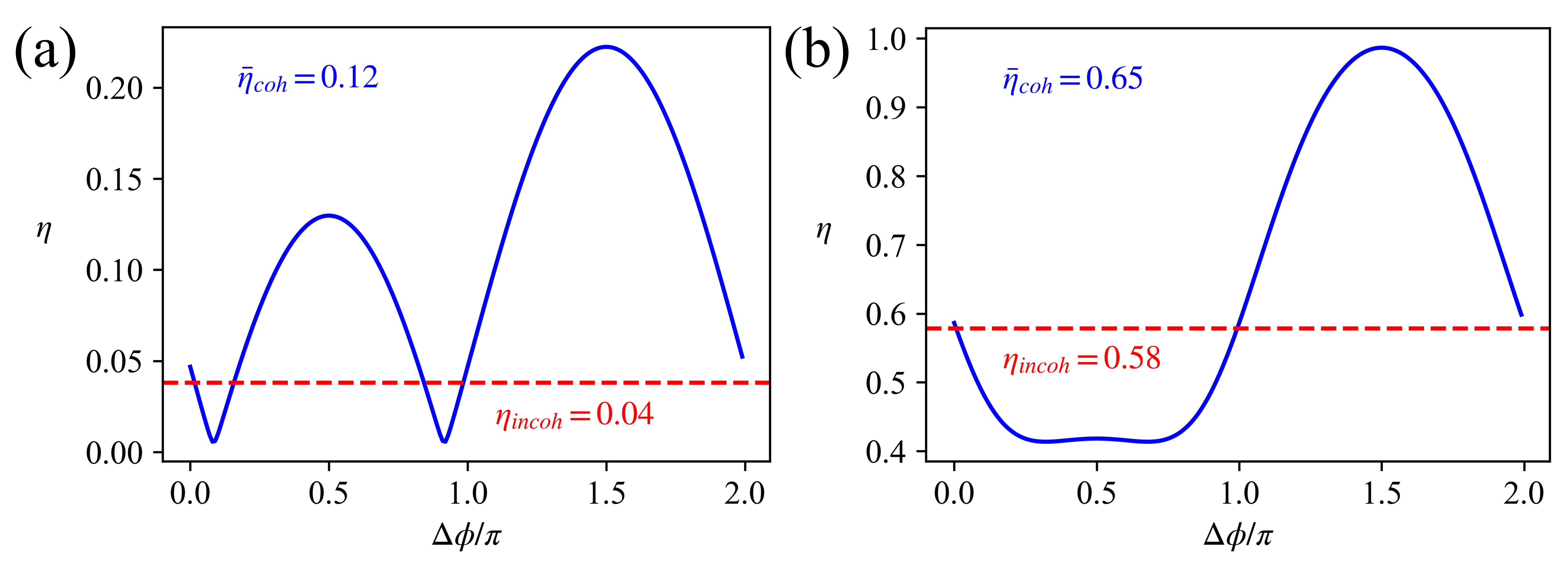}
\caption{Mismatch index $\eta$ for exciting rings \#3 and \#4 with a coherent input signal having a relative phase of $\Delta\phi$, as predicted by TCMT, based on the experimental model parameters. The analysis is performed considering two scenarios for the intrinsic loss of the rings: (a) $l_o=0$ and (b) $l_o=0.56$ GHz.}
\label{fig:eta_34}
\end{figure}
\end{center}
\begin{center}
\begin{figure*}
\includegraphics[width=0.9\textwidth]{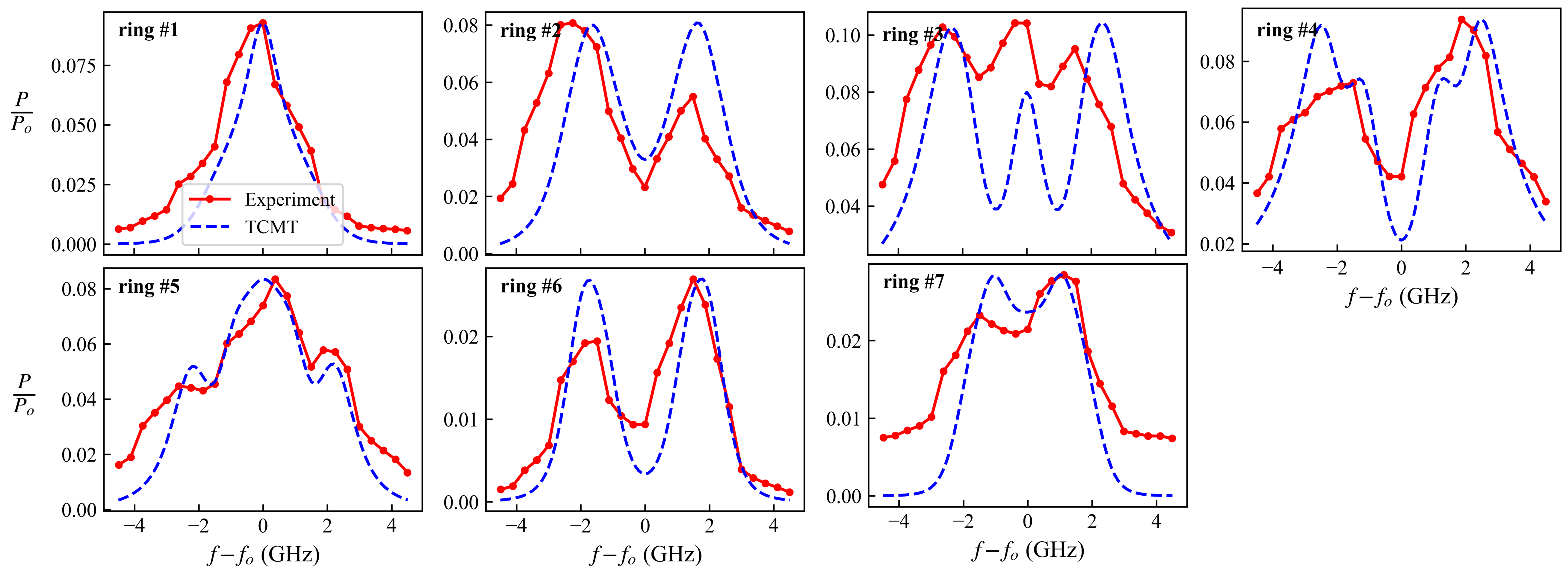}
\caption{Power spectrum of the ring resonators obtained from experimental measurements (red line) and TCMT analysis (dashed blue line) for incoherent excitation of rings \#3 and \#4. Incoherent excitation is realized by averaging 50 coherent excitations with randomly varied relative phases. The experimental data is normalized so that the peak power in each spectrum aligns with the maximum power predicted by TCMT.}
\label{fig:incoh_34_spec}
\end{figure*}
\end{center}

\begin{center}
\begin{figure}
\includegraphics[width=2.3in]{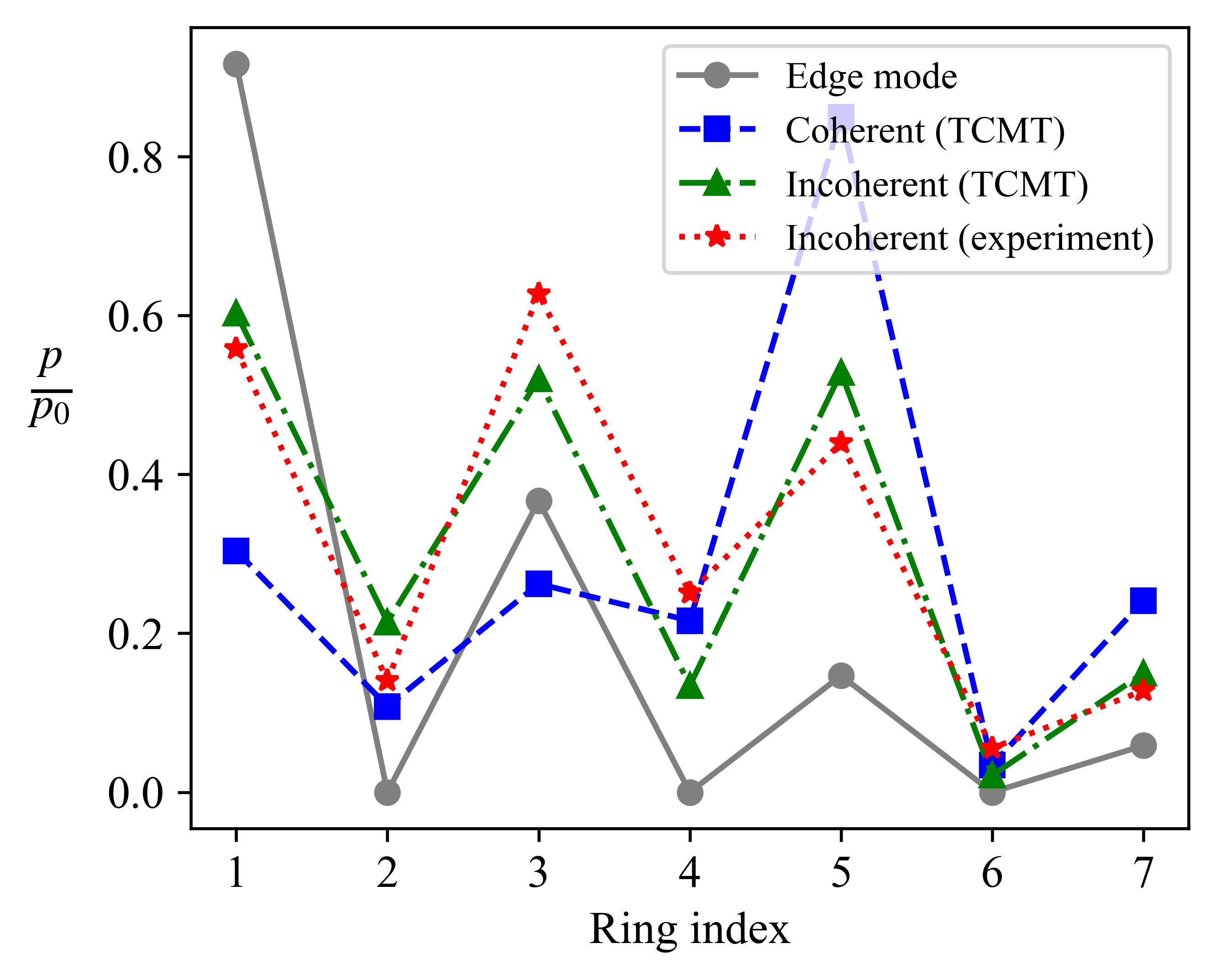}
\caption{The spatial power distribution of the edge state (gray line) is compared to the system's spatial power distribution after exciting the system with a coherent input at frequency $f=f_o$, coupled to rings \#3 and \#4, as predicted by TCMT (dashed blue line). Additionally, the spatial power distribution for an incoherent input signal is shown, with experimental results (dotted red line) and TCMT predictions (dashed-dotted green line) presented for comparison.}
\label{fig:incoh_34_power}
\end{figure}
\end{center}
Next, we examine a less intuitive scenario where an incoherent multiport input signal can enhance mode excitation even when the input has significant overlap with multiple modes. For instance, consider an input signal exciting rings \#3 and \#4. The input port of ring \#4 has no direct overlap with the edge mode but strongly overlaps with modes \#1, 3, 5, and 7. Similarly, while the input port of ring \#3 has considerable overlap with the edge mode, it also strongly overlaps with other modes, such as modes \#1, 2, 6, and 7. Consequently, the input signal exciting rings \#3 and \#4 not only excites the edge mode but also leads to significant excitation of other modes.

Figure \ref{fig:eta_34}(a) presents the mismatch index for a two-port excitation at rings \#3 and \#4 as a function of the relative phase $\Delta\phi$, assuming zero intrinsic loss in the ring resonators. This assumption is made to emphasize the high sensitivity of the system to phase adjustments. Unlike the previous case, the mismatch index curve is no longer flat near its minimum, indicating that incoherent excitation outperforms coherent excitation when the input phase is not precisely tuned.
The sensitivity of mode excitation to the relative phase is more pronounced in this two-port excitation scheme compared to the case of exciting rings \#1 and \#3 which is due to the stronger coupling of the input ports at rings \#3 and \#4 with multiple modes beyond the edge state.
Figure \ref{fig:eta_34}(b) shows the mismatch index as a function of the relative phase, but considers the intrinsic loss present in the ring resonators used in our experiment.
While intrinsic loss reduces the impact of phase sensitivity, the results still demonstrate that incoherent excitation achieves better mode selectivity compared to randomly phased coherent excitation, even in a low-$Q$ system.

The spatial optical power distribution for the incoherent input exciting rings \#3 and \#4 is obtained by first averaging the power spectra over 50 coherent input signals with randomly varying relative phases, as shown in Fig.~\ref{fig:incoh_34_spec} (red solid line for experimental data and dashed blue line for TCMT predictions), and then extracting the spatial power distribution at $f=f_o$. The resulting distribution is depicted in Fig.~\ref{fig:incoh_34_power}, yielding a mismatch index of $\eta_{exp}^{incoh} = 0.61$ from experimental measurements (red dotted line) and $\eta_{TCMT}^{incoh} = 0.58$ from TCMT analysis (green dashed-dotted line). These values confirm that incoherent excitation effectively enhances the selective excitation of the edge mode.
To further highlight the impact of phase tuning in coherent excitation, Fig.~\ref{fig:incoh_34_power} also plots the spatial power distribution for a coherent input with a relative phase of $\Delta\phi = 3\pi/2$ (dashed blue line), resulting in a significantly higher mismatch index of $\eta_{TCMT}^{coh} = 0.99$. This demonstrates that when the relative phase deviates from the optimal value, the excitation efficiency of the edge mode is severely degraded, leading to substantial power distribution among other modes. 
These results further reinforce the key observation that incoherent excitation offers a more robust approach in scenarios where phase sensitivity is high. Even in a system with non-negligible intrinsic loss, such as the one used in our experiment, incoherent excitation consistently outperforms random-phase coherent excitation, leading to a more efficient excitation of the edge mode.

Having established the excitation behavior in the ideal system, we now examine the robustness of the excited topological mode under incoherent excitation. To this end, we introduce deliberate disorder in the coupling coefficients between the ring resonators, following a Gaussian distribution with a standard deviation of $\sigma_{1,2}=0.1\times t_{1,2}$. We generate 500 distinct disorder realizations of the system and excite each configuration with incoherent inputs for two representative cases. In Case 1, the edge mode is excited through input ports coupled to rings \#1 and \#3, while in Case 2, a bulk mode (mode \#2) is excited through input ports coupled to rings \#2 and \#4 (see Appendix~\ref{appendix_bulk_excitation}). Figures~\ref{fig:rnd}(a) and~(b) show the normalized steady-state power distributions along the ring array for Cases 1 and 2, respectively. As evident from these figures, the incoherent input continues to selectively excite the desired modes, yielding ensemble-averaged mismatch indices of $\langle\eta_{incoh}^{edge}\rangle = 0.276$ and $\langle\eta_{incoh}^{bulk}\rangle = 0.267$ for the edge and bulk modes, respectively, where $\langle \cdot\rangle$ denotes averaging over all disorder realizations. Notably, for both cases, the mean mismatch index remains close to that of the corresponding ideal (disorder-free) systems, with standard deviations of $\sigma_{\eta}^{edge} = 0.01$ and $\sigma_{\eta}^{bulk}=0.008$. It is also worth emphasizing that the edge-state excitation exhibits slightly higher robustness compared to the bulk-state excitation. This is reflected in the smaller average standard deviation of the normalized power distribution, $\sigma_p^{edge}=0.002$, which is approximately two times lower than that of the bulk mode excitation, $\sigma_p^{bulk}=0.004$. These results demonstrate that incoherent excitation provides statistical stability against coupling disorder and preserves the inherent robustness of the topological edge mode.

\begin{figure}
    \centering
    \includegraphics[width=3.4in]{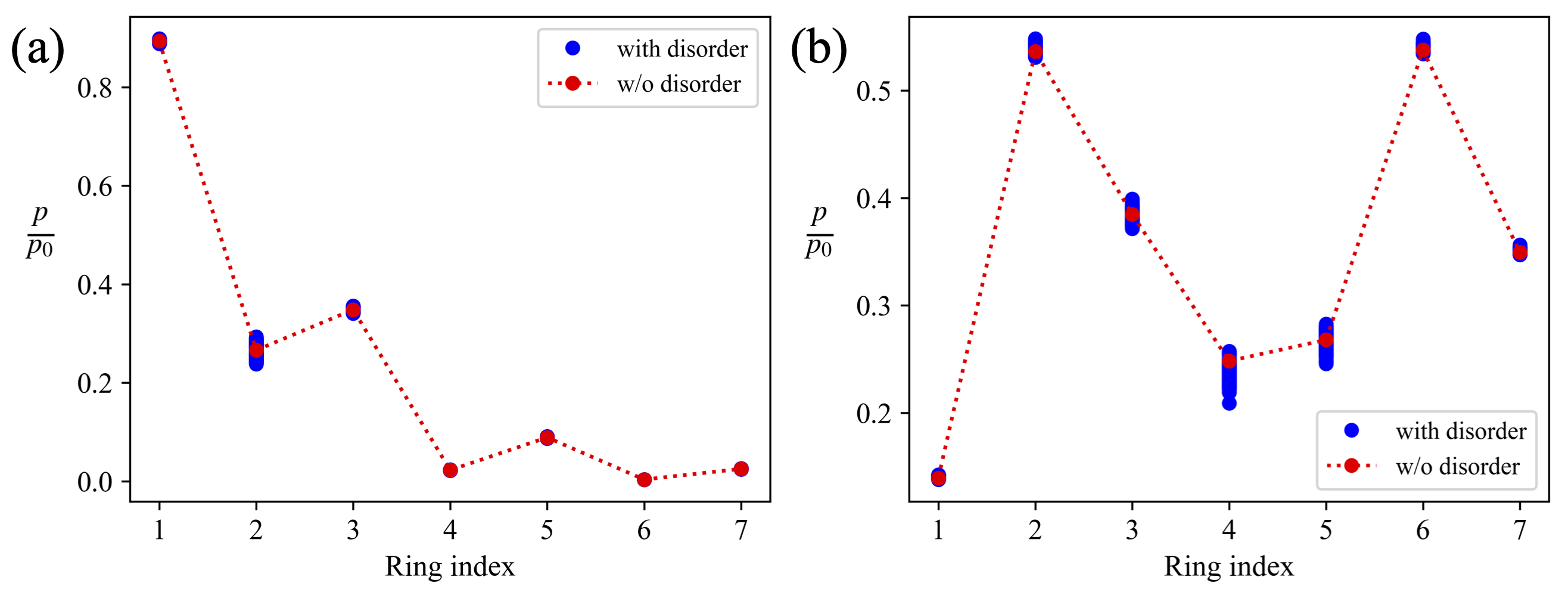}
    \caption{Spatial power distribution along the ring resonators for (a) edge-mode excitation, and (b) bulk-mode (mode \#2) excitation through incoherent input, for 500 disorder realizations with coupling fluctuations $\sigma_{1,2}=0.1\times t_{1,2}$.}
    \label{fig:rnd}
\end{figure}

Finally, it is worth mentioning that the proposed approach can also be extended to systems operating near exceptional points. By considering the resolvent (Green’s) operator formalism for such systems, as discussed in Ref. \cite{Hashemi2022NatComm}, it can be seen that the non-zero decay rate of the modes still limits the excitation efficiency. In this regime, employing a multiport input configuration can enhance the overlap between the excitation field and the desired mode. Similar to passive systems, the excitation efficiency in this case remains sensitive to the relative phase of the input ports, and incoherent excitation can provide improved robustness compared to coherent inputs with imperfect phase control.

\section{conclusion}
In this study, we investigated the role of incoherent excitation in selectively exciting modes in non-Hermitian resonant systems. By implementing a topological non-Hermitian SSH model using a photonic platform with microring resonators, we demonstrated that incoherent input signals can enhance excitation efficiency compared to coherent excitation with an arbitrary phase, but not relative to optimally phase-matched coherent excitation.
For cases where the input ports are primarily coupled to a single mode, such as the excitation of rings \#1 and \#3, coherent excitation with a properly tuned relative phase (close to $\pi$) effectively excites the edge mode while suppressing other modes. However, we showed that incoherent excitation still provides a more robust excitation scheme, as it eliminates the sensitivity to phase fluctuations and achieves a lower mismatch index on average.
In scenarios where the input ports have significant overlap with multiple modes, such as the excitation of rings \#3 and \#4, we found that phase sensitivity is more pronounced. When the relative phase deviates from its optimal value, mode selectivity is severely degraded, leading to a high mismatch index. Our results show that incoherent excitation outperforms phase-mismatched coherent excitation in such cases, ensuring a more efficient and stable excitation of the edge mode.
Furthermore, we analyzed the impact of intrinsic losses in our experimental system, which consists of ring resonators with a moderate $Q$ factor of $Q \approx 4 \times 10^4$. Even in the presence of loss, the advantage of incoherent excitation persists, reinforcing its potential as a practical strategy for enhancing mode selectivity in non-Hermitian photonic systems.
Overall, our findings highlight the fundamental role of incoherence in mitigating phase sensitivity and improving excitation efficiency in non-Hermitian resonant systems. These results could be relevant for various applications of photonic topological insulators and non-Hermitian systems, where precise mode control is essential.

\section*{Acknowledgements}
A.B.-R. acknowledges support by the NSF award number 2328993. A P.-L. is supported by MURI grant from Air Force Research Office (programmable systems with non-Hermitian quantum dynamics: FA9550-21-1-0202)

\section*{DATA AVAILABILITY}
The data that support the findings of this article are available from the authors
upon request.

\appendix
\section{Coherent excitation of non-Hermitian resonant systems}\label{appendix_transient}
\begin{figure}
    \centering
    \includegraphics[width=3in]{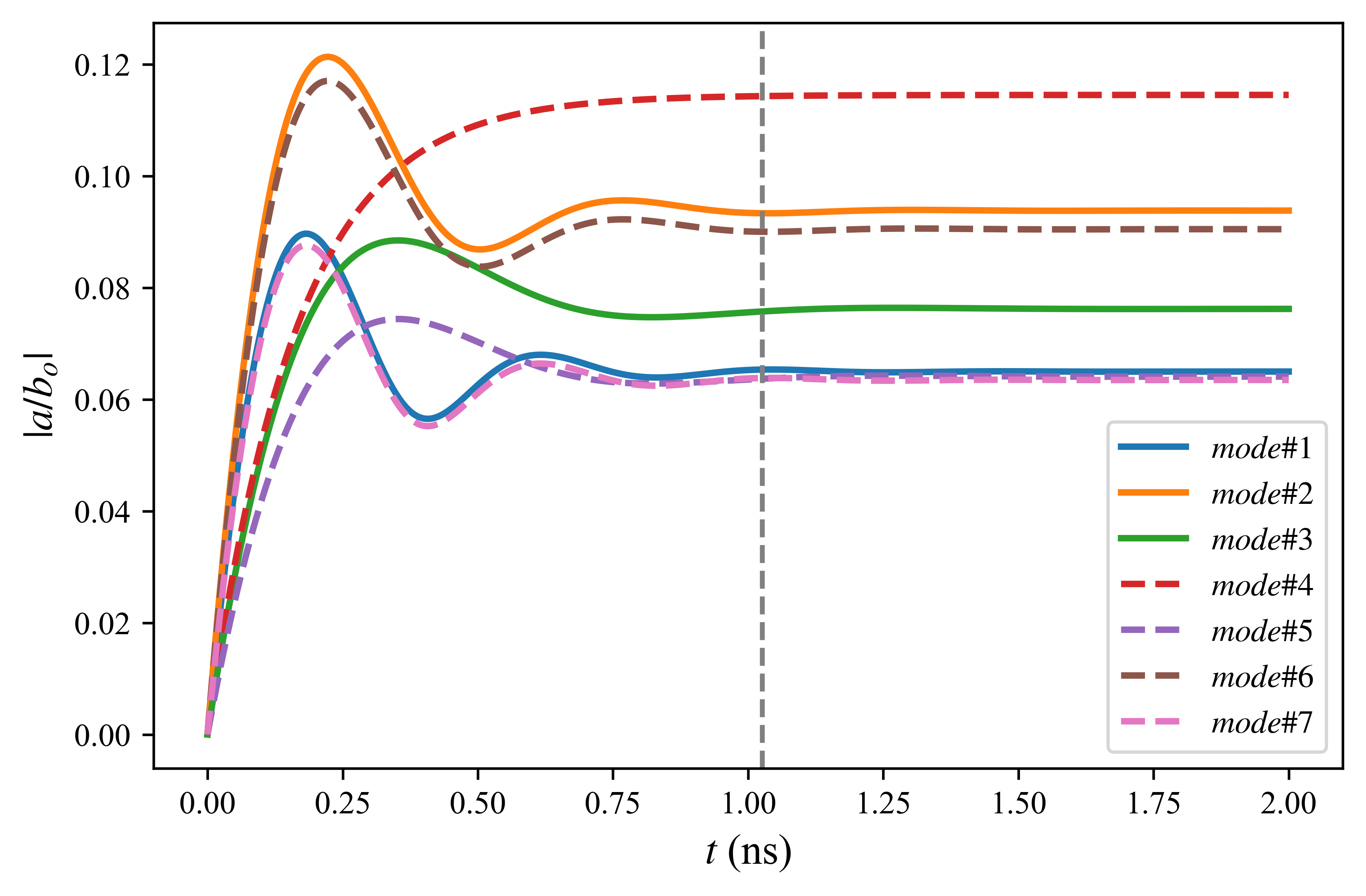}
    \caption{Temporal evolution of the system's mode for a coherent excitation of the experimental platform shown in Fig.~\ref{fig:mesh}(b) obtained through TCMT analysis. The vertical gray dashed line shows the system's constant time $\tau_s= 1.02$~(ns).}
    \label{fig:transient}
\end{figure}

The formal solution of Eq.~(\ref{eq:tcmt}) is given by:
\begin{align}\label{eq:TCMT_sol_general}
    \ket{a(t)}=U(t)\ket{a_o}+\int_0^t U(t-t')\hat{\Gamma}\ket{b(t')}dt',
\end{align}
where $\ket{a_o}\equiv \ket{a(t=0)}$, and $U(t)\equiv e^{-i\hat{H}t}$ is the system evolution operator.
For a coherent input signal $\ket{b(t)}=\ket{b_o}e^{-i\omega t}$, Eq.~(\ref{eq:TCMT_sol_general}) 
can be further simplified by expanding the evolution operator in the eigenbasis of the system Hamiltonian resulting in the following solution:
\begin{align}
    \label{eq:TCMT_sol}
    \ket{a(t)} = & \sum_{n} \ket{\psi_n^r}\braket{\psi_n^l|a_o}e^{-i\Omega_n t} \nonumber\\
    & - \sum_n \frac{\ket{\psi_n^r}\braket{\psi_n^l|i\hat{\Gamma}|b_o}}{\omega-\Omega_n}\left(e^{-i\Omega_n t} - e^{-i\omega t} \right).
\end{align}
For a passive system, the intrinsic losses cause all eigenfrequencies to possess negative imaginary parts, $\mathrm{Im}(\Omega_n) < 0$, whose absolute values correspond to the decay rates of the respective modes. Consequently, the first and second terms in Eq.~(\ref{eq:TCMT_sol}) represent transient contributions that decay exponentially and vanish after a characteristic system time $t > \tau_s$, where $\tau_s \equiv 2\pi/\Gamma$ and $\Gamma \equiv \text{Max}(|\mathrm{Im}(\Omega_n)|)$ denotes the largest modal decay rate. Beyond this time, the system reaches a steady state, which is described by the third term in Eq.~(\ref{eq:TCMT_sol}).\\
For our experimental platform shown in Fig.~\ref{fig:mesh}(b), the temporal evolution of the modes under coherent excitation through rings~\#1 and~\#3 is presented in Fig.~\ref{fig:transient}, illustrating that the system initially undergoes a transient evolution lasting approximately $\tau_s= 1.02$~ns, after which it reaches a steady-state response.

\section{Time domain framework and master equation}\label{appendix_master_eq}
\begin{figure}
    \centering
    \includegraphics[width=0.9\linewidth]{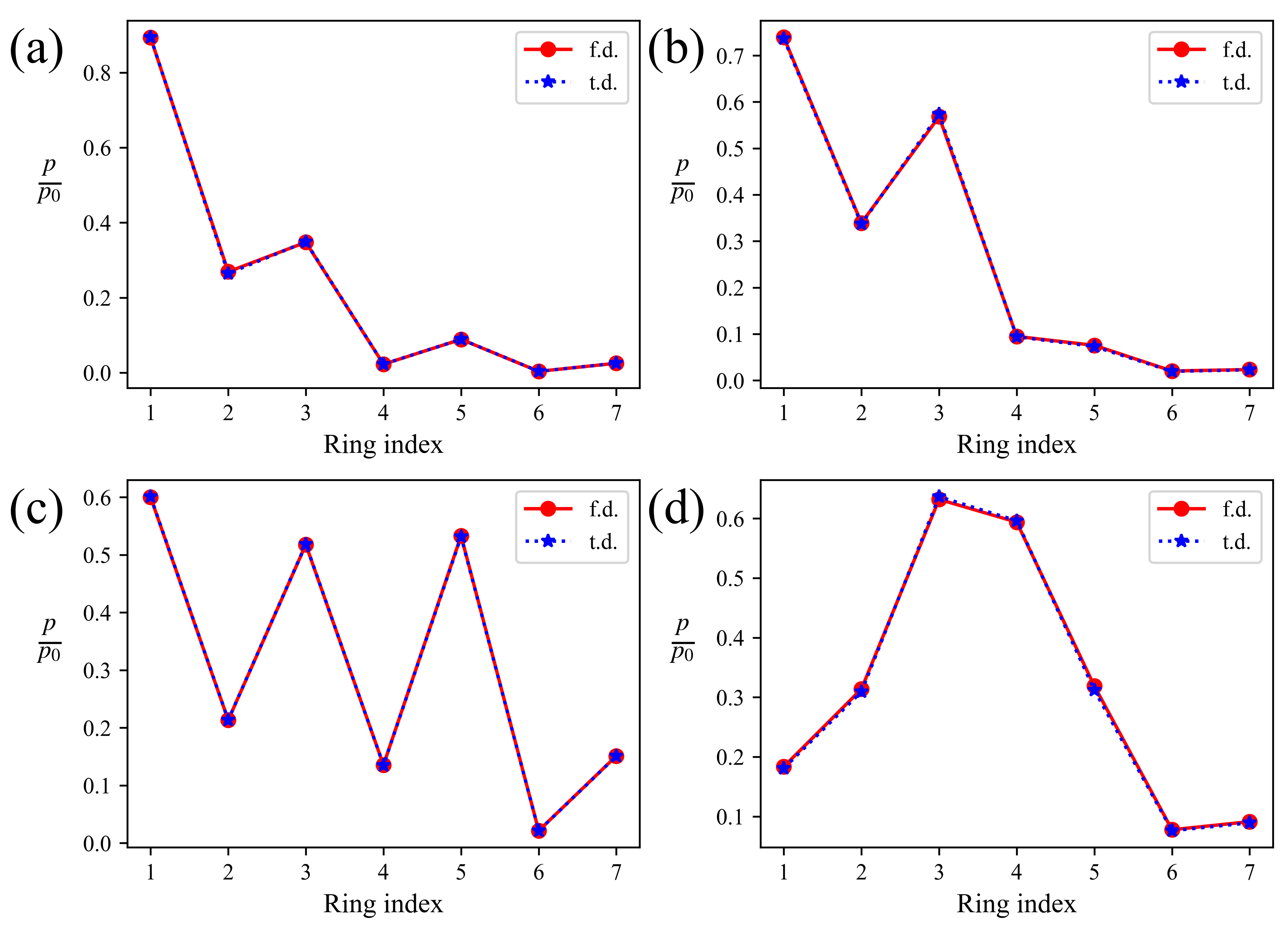}
    \caption{Spatial power distribution along the ring resonators under incoherent excitation. (a),(b) Excitation of rings \#1 and \#3 with narrowband input at $\omega_o$ and broadband (white-light) input, respectively. (c),(d) Same as (a),(b) for excitation of rings \#3 and \#4. Results from the time-domain master equation (blue) and the frequency-domain resolvent approach (red) are in excellent agreement.}
    \label{fig:time_domain}
\end{figure}

The state of the non-Hermitian resonant system driven by an incoherent input signal is described by the system coherence matrix $\hat{\rho}(t)\equiv \braket{\ket{a}\bra{a}}$. The master equation governing the time evolution of $\hat{\rho}(t)$ is obtained by taking its time derivative and employing Eq.~(\ref{eq:tcmt}) together with its Hermitian conjugate to eliminate the derivatives of $\ket{a}$ and $\bra{a}$. This yields
\begin{align}\label{eq:master_derivation}
\frac{d\hat{\rho}}{dt}=-i\left(\hat{H}\hat{\rho}-\hat{\rho}\hat{H}^\dagger\right) + \hat{\Gamma}\braket{\ket{b}\bra{a}} + \braket{\ket{a}\bra{b}}\hat{\Gamma}^\dagger.
\end{align}
Under the assumption of incoherent excitation, where the coherence time $\tau_c$ exceeds the system time constant $\tau_s$ (i.e., $\tau_c > \tau_s$), Eq.~(\ref{eq:TCMT_sol_general}) and its Hermitian conjugate can be invoked to substitute for $\ket{a}$ and $\bra{a}$ in the third and fourth terms of Eq.~(\ref{eq:master_derivation}). This procedure leads to Eq.~(\ref{eq:master}).\\
The solution of the master equation depends on the statistical properties of the input field, characterized by the coherence matrix $\hat{\rho}_s(t,t')\equiv\braket{\ket{b(t)}\bra{b(t')}}$, where $\ket{b(t)}$ denotes the input field vector driving the system. Assuming that the input coherence matrix satisfies the conditions of the Wiener–Khinchin theorem~\cite{statisticalopticsbook}, it can be expressed as
\begin{align}\label{eq:power_spec}
\hat{\rho}_s(t,t')=\frac{1}{2\pi}\int_{-\infty}^{+\infty}\hat{S}(\omega) e^{-i\omega (t-t')}d\omega,
\end{align}
where $\hat{S}(\omega)$ denotes the power spectral density of the input signal.\\
For resonant systems, one is typically interested in the response to excitations with frequencies close to a system resonance. In this regime, the power spectral density of the input signal can be approximated as $\hat{S}(\omega)=2\pi\hat{S}_o\delta(\omega-\omega_o)$, where $\omega_o$ denotes the resonance frequency and $\hat{S}_o$ is a constant matrix. For the spatially incoherent input considered in this work, $\hat{S}_o$ is diagonal, thereby precluding correlations between fields at different input ports.
For such an ideally narrowband input field, the coherence matrix reduces to $\hat{\rho}_s(t,t')=\hat{S}_o e^{-i\omega_o(t-t')}$. Substituting this expression into Eq.~(\ref{eq:master_source}) and expanding the time-evolution operator in the basis of the Hamiltonian eigenvectors yields
\begin{align}
\hat{D}=i\sum_n\frac{\ket{\psi_n^r}\bra{\psi_n^l}}{\omega_o-\Omega_n}\hat{\Gamma}\hat{S}_o\hat{\Gamma}^\dagger\left(1-e^{-i(\Omega_n-\omega_o)t}\right)+h.c.
\end{align}
For the lossy resonant system considered here, $\mathrm{Im}(\Omega_n)\neq 0$ for all modes, ensuring that the transient terms decay in the long-time limit. Consequently, one obtains $\hat{D}(t\rightarrow\infty)=i\hat{G}(\omega_o)\hat{\Gamma}\hat{S}_o\hat{\Gamma}^\dagger+h.c.$\\
Another extreme case corresponds to a white-light input, i.e., a broadband excitation with a flat spectral profile uniformly spanning all frequencies. In this regime, the input coherence matrix is given by $\hat{\rho}_s(t,t')=\hat{S}_o\delta(t-t')$, which follows from Eq.~(\ref{eq:power_spec}) upon assuming a frequency-independent power spectral density $\hat{S}(\omega)=\hat{S}_o$.
Substitution of this expression into Eq.~(\ref{eq:master_source}) yields $\hat{D}=\hat{\Gamma}\hat{S}_o\hat{\Gamma}^\dagger$. In performing the time integration involving the Dirac delta function, the Stratonovich midpoint rule has been adopted \cite{SDEbook}, ensuring a consistent treatment of the singular correlation function.\\
To demonstrate the equivalence between the time-domain framework and the frequency-domain approach employed in this work, Fig.~\ref{fig:time_domain} presents the spatial power distribution of the system obtained using the time-domain master equation (blue) and the frequency-domain resolvent formalism (red). Results are shown for excitation of rings 1 and 3 [Fig.~\ref{fig:time_domain}(a),(b)] and rings 3 and 4 [Fig.~\ref{fig:time_domain}(c),(d)], for both narrowband and broadband input signals, respectively. Good agreement between the two approaches is observed.

\section{Phase instability in multi-Port excitation}\label{appendix_phase_flactuation}
\begin{figure}
    \centering
    \includegraphics[width=0.9\linewidth]{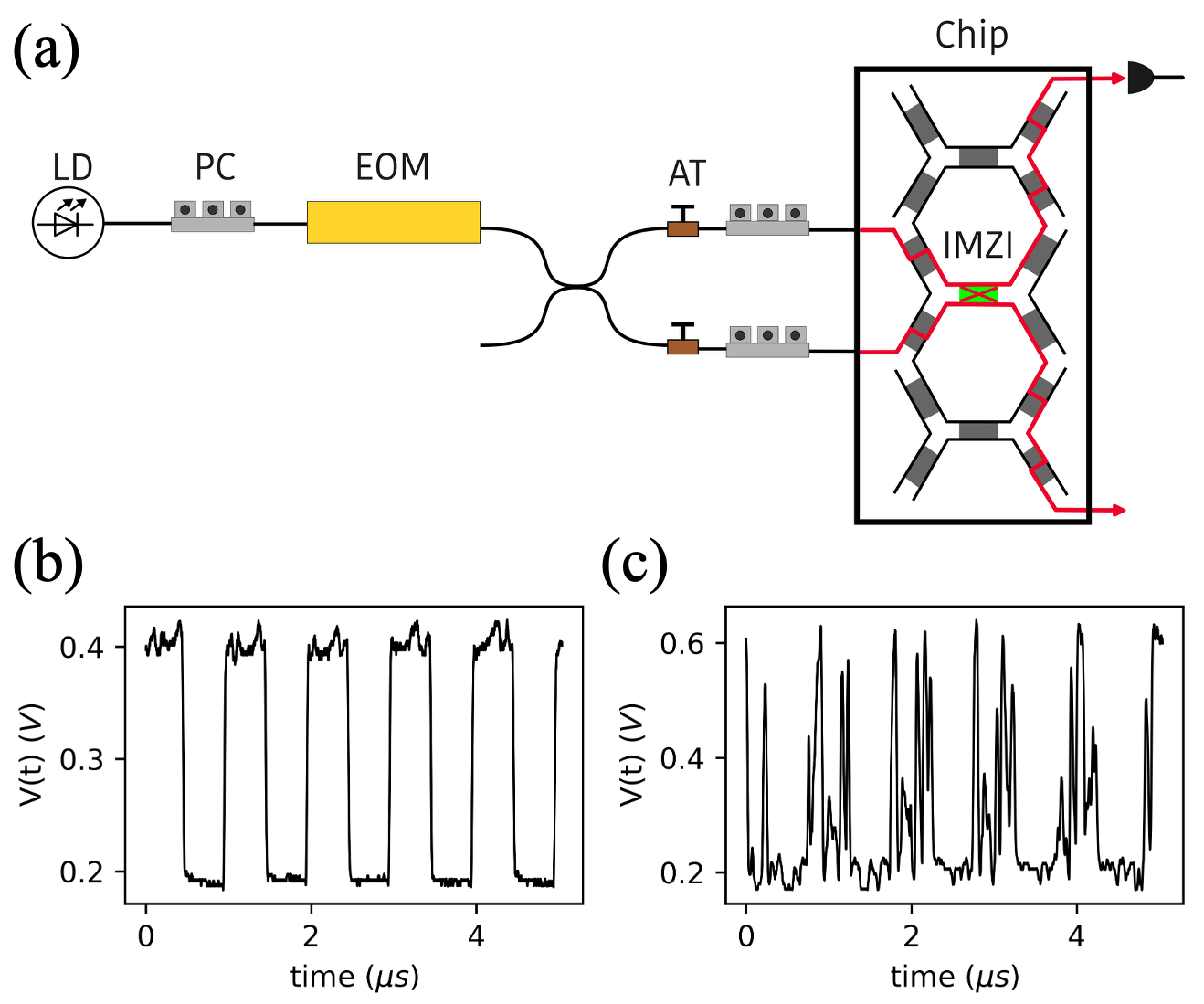}
    \caption{(a) Schematic of the experimental setup used to verify input phase instability. Interference occurs in the on-chip 3-dB MZI, labeled as IMZI and highlighted in green. LD: laser diode; PC: polarization controller; TA: tunable attenuator; EOM: electro-optic modulator. (b) Modulated input signal at one output of the beam splitter with a modulation frequency of 1~MHz. (c) Corresponding output signal from one port of the on-chip MZI. The strong degradation of the modulation in (c) indicates fluctuations of the relative phase between the two inputs.}
    \label{fig:phase_flactuation}
\end{figure}

While integrated photonic platforms can, in principle, support multi-port excitation with well-defined relative phases, achieving such control typically requires careful system design. For example, a coherent laser source can be coupled into the chip and subsequently split into multiple on-chip paths, each routed to a distinct input port. By incorporating tunable phase shifters along these paths, it is possible to precisely control and stabilize the relative phases of the input fields.
In the absence of such phase-stabilization mechanisms, however, the relative phases between different input channels are generally not well controlled. For instance, if light is delivered to multiple input ports via separate optical fibers without active phase locking, environmental perturbations (e.g., thermal fluctuations or mechanical vibrations) introduce random phase variations between the inputs.
To illustrate this effect, we consider the simple configuration shown in Fig.~\ref{fig:phase_flactuation}(a). A laser source is first passed through an amplitude modulator, after which the modulated signal is split by a fiber-optic beam splitter. The two outputs are then independently coupled into the chip and fed into the two input ports of an on-chip 3 dB MZI. Figure~\ref{fig:phase_flactuation}(b) shows the modulated input signal at one output of the beam splitter, while Fig.~\ref{fig:phase_flactuation}(c) shows the corresponding output signal from one port of the MZI.
In the ideal case of a stable relative phase between the two MZI inputs, the output would preserve the modulation profile. However, as observed in Fig.~\ref{fig:phase_flactuation}(c), the modulation is significantly degraded, indicating substantial phase fluctuations between the input channels. In fact, the relative phase spans the full $0$ to $2\pi$ range, leading to inversion of the interference from maxima to minima. This behavior provides direct evidence of phase instability in the absence of active control. Supplementary videos 1 and 2 further illustrate the temporal fluctuations of both the input and interfered signals, respectively.

\section{Incoherent excitation of a bulk mode}\label{appendix_bulk_excitation}
\begin{figure}[t]
    \centering
    \includegraphics[width=3.4in]{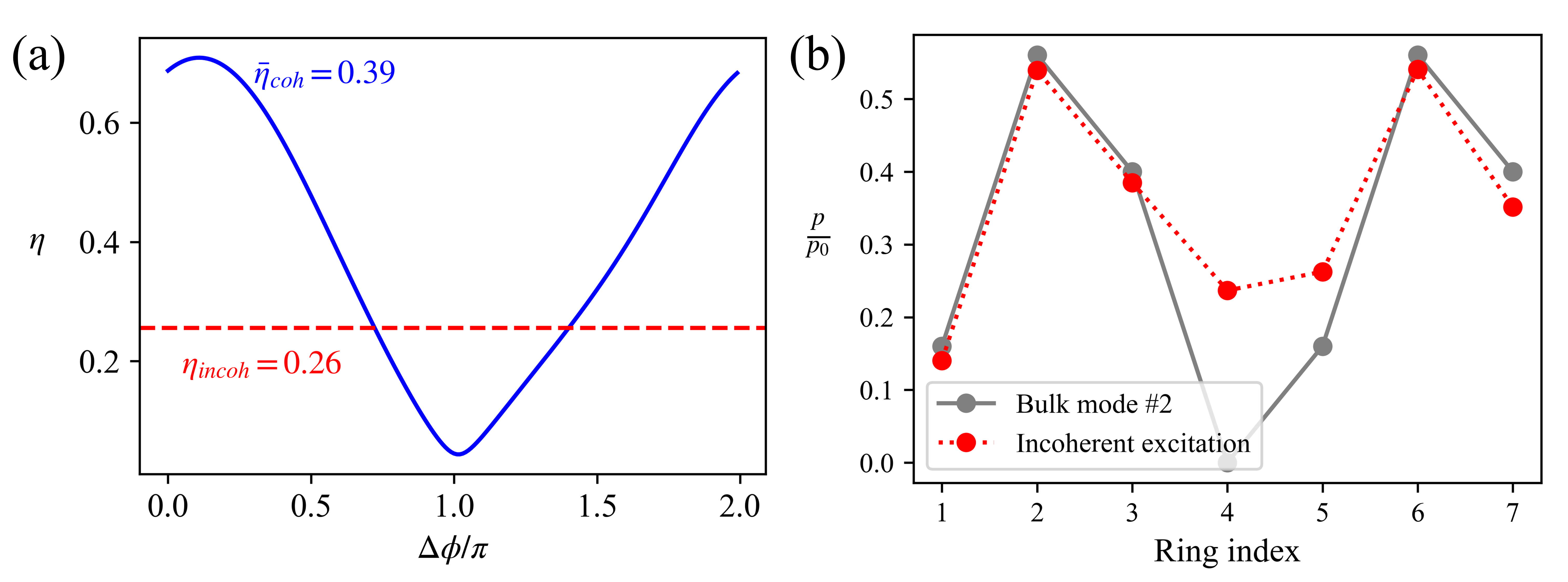}
    \caption{(a) Mismatch index $\eta$ for exciting rings \#2 and \#6 with a coherent input signal having a relative phase of $\Delta\phi$, as predicted by TCMT, based on the experimental model parameters. (b) Spatial power distribution for the incoherent excitation (red line) compared with the eigenmode profile of bulk mode~\#2 (gray line)}
    \label{fig:bulk}
\end{figure}

In this section, we analyze the proposed incoherent excitation scheme for selectively exciting a bulk mode of the system shown in Fig.~\ref{fig:mesh}, specifically mode~\#2. As illustrated in Fig.~\ref{fig:mesh}(e), the input ports coupled to rings~\#2 and~\#6 exhibit strong spatial overlap with this mode; therefore, these ports are chosen for excitation. Figure~\ref{fig:bulk}(a) presents the mismatch index as a function of the relative phase between the input signals, demonstrating the improved efficiency of the incoherent excitation, with $\eta_{incoh} = 0.26$, compared to the averaged coherent excitation, $\bar{\eta}_{coh} = 0.39$. The corresponding spatial power distribution for the incoherent excitation is shown in Fig.~\ref{fig:bulk}(b) (red line), which closely matches that of bulk mode~\#2 (gray line), confirming the expected excitation behavior. It is worth noting that, for a fixed number of input ports, spatially localized modes (such as the topological edge mode) can generally be excited more efficiently than modes that are extended across the entire system (such as bulk modes), regardless of whether coherent or incoherent excitation is employed. As a result, using only two input ports to excite bulk modes is inherently less efficient than exciting a localized topological mode.

\section{Numerical simulations}\label{appendix_numerical_simulations}
To model incoherent excitation in the frequency-domain framework, the input field is treated as a statistical ensemble rather than a single phase-defined waveform. Specifically, the amplitudes injected through the input ports are kept fixed while their relative phases are randomized with a uniform distribution within $[0,2\pi)$ range, representing a spatially incoherent input. 
For example, for excitation of rings \#1 and \#3 with an incoherent input at frequency $\omega_o$ [Fig.~\ref{fig:13_power}(b)], the input field is taken as $\ket{B(\omega)}=[e^{i\phi_1},0,e^{i\phi_2},0,0,0,0]^T\delta(\omega-\omega_o)$, where the phases $\phi_{1,2}$ are treated as independent random variables. In contrast, for coherent excitation [Fig.~\ref{fig:13_power}(a)], these phases are fixed. For the spectra shown in Fig.~\ref{fig:incoh_13_spec} (Fig.~\ref{fig:coh_13_spec}), the input signal is given by $\ket{B(\omega)}=[e^{i\phi_1},0,e^{i\phi_2},0,0,0,0]^T$, with random (fixed) phases for incoherent (coherent) excitation.
For each realization of the input phases, the system response is computed using the deterministic frequency-domain solution of the coupled-mode equations given in Eq.\ref{eq:freq_response}. The steady-state observables under incoherent driving—here, the intracavity powers—are obtained by ensemble averaging over 1000 independent phase realizations. Convergence is verified by increasing the number of phase realizations until the ensemble-averaged observables become independent of the sample size.
In this sense, the simulation implements a Monte Carlo ensemble method \cite{MonteCarloSimulationBool}, where stochasticity enters only through the input statistics while the system evolution itself remains deterministic. The approach is equivalent to propagating the second-order statistical properties of the input through the linear system response and provides a computationally straightforward alternative to stochastic differential-equation methods. The averaged quantities therefore represent experimentally measurable powers under incoherent excitation. From the resulting intracavity powers under incoherent and coherent excitation, the mismatch index is computed using Eq.~(\ref{eq:eta}), as shown in Fig.~\ref{fig:eta_13}.

\bibliography{References.bib}

@article{Koch2022PysRevRes,
  title = {Quantum non-Hermitian topological sensors},
  author = {Koch, Florian and Budich, Jan Carl},
  journal = {Phys. Rev. Res.},
  volume = {4},
  issue = {1},
  pages = {013113},
  numpages = {8},
  year = {2022},
  month = {Feb},
  publisher = {American Physical Society},
  doi = {10.1103/PhysRevResearch.4.013113},
  url = {https://link.aps.org/doi/10.1103/PhysRevResearch.4.013113}
}

@Article{McDonald2020NatComm,
author={McDonald, Alexander
and Clerk, Aashish A.},
title={Exponentially-enhanced quantum sensing with non-Hermitian lattice dynamics},
journal={Nature Communications},
year={2020},
month={Oct},
day={23},
volume={11},
number={1},
pages={5382},
issn={2041-1723},
doi={10.1038/s41467-020-19090-4},
url={https://doi.org/10.1038/s41467-020-19090-4}
}

@Article{Ozdemir2019NatMat,
author={{\"O}zdemir, {\c{S}} K.
and Rotter, S.
and Nori, F.
and Yang, L.},
title={Parity--time symmetry and exceptional points in photonics},
journal={Nature Materials},
year={2019},
month={Aug},
day={01},
volume={18},
number={8},
pages={783-798},
issn={1476-4660},
doi={10.1038/s41563-019-0304-9},
url={https://doi.org/10.1038/s41563-019-0304-9}
}

@ARTICLE{Wonjoo2004IEEE,
  author={Wonjoo Suh and Zheng Wang and Shanhui Fan},
  journal={IEEE Journal of Quantum Electronics}, 
  title={Temporal coupled-mode theory and the presence of non-orthogonal modes in lossless multimode cavities}, 
  year={2004},
  volume={40},
  number={10},
  pages={1511-1518},
  doi={10.1109/JQE.2004.834773}}

@article{Fan2003josaA,
author = {Shanhui Fan and Wonjoo Suh and J. D. Joannopoulos},
journal = {J. Opt. Soc. Am. A},
number = {3},
pages = {569--572},
publisher = {Optica Publishing Group},
title = {Temporal coupled-mode theory for the Fano resonance in optical resonators},
volume = {20},
month = {Mar},
year = {2003},
url = {https://opg.optica.org/josaa/abstract.cfm?URI=josaa-20-3-569},
doi = {10.1364/JOSAA.20.000569}
}

@Article{Hashemi2022NatComm,
author={Hashemi, A.
and Busch, K.
and Christodoulides, D. N.
and Ozdemir, S. K.
and El-Ganainy, R.},
title={Linear response theory of open systems with exceptional points},
journal={Nature Communications},
year={2022},
month={Jun},
day={07},
volume={13},
number={1},
pages={3281},
issn={2041-1723},
doi={10.1038/s41467-022-30715-8},
url={https://doi.org/10.1038/s41467-022-30715-8}
}

@article{SSH1979PhysRevLett,
  title = {Solitons in Polyacetylene},
  author = {Su, W. P. and Schrieffer, J. R. and Heeger, A. J.},
  journal = {Phys. Rev. Lett.},
  volume = {42},
  issue = {25},
  pages = {1698--1701},
  numpages = {0},
  year = {1979},
  month = {Jun},
  publisher = {American Physical Society},
  doi = {10.1103/PhysRevLett.42.1698},
  url = {https://link.aps.org/doi/10.1103/PhysRevLett.42.1698}
}

@article{Lieu2018PhysRevB,
  title = {Topological phases in the non-Hermitian Su-Schrieffer-Heeger model},
  author = {Lieu, Simon},
  journal = {Phys. Rev. B},
  volume = {97},
  issue = {4},
  pages = {045106},
  numpages = {7},
  year = {2018},
  month = {Jan},
  publisher = {American Physical Society},
  doi = {10.1103/PhysRevB.97.045106},
  url = {https://link.aps.org/doi/10.1103/PhysRevB.97.045106}
}

@Article{Hashemi2025NatMat,
author={Hashemi, Amin
and Pereira, Elizabeth Louis
and Li, Hongwei
and Lado, Jose L.
and Blanco-Redondo, Andrea},
title={Observation of non-Hermitian topology from optical loss modulation},
journal={Nature Materials},
year={2025},
month={Jul},
day={23},
issn={1476-4660},
doi={10.1038/s41563-025-02278-8},
url={https://doi.org/10.1038/s41563-025-02278-8}
}

@article{Meng2024APL,
    author = {Meng, Haiyu and Ang, Yee Sin and Lee, Ching Hua},
    title = {Exceptional points in non-Hermitian systems: Applications and recent developments},
    journal = {Applied Physics Letters},
    volume = {124},
    number = {6},
    pages = {060502},
    year = {2024},
    month = {02},
    issn = {0003-6951},
    doi = {10.1063/5.0183826},
    url = {https://doi.org/10.1063/5.0183826}
}

@article{Takata2018PysRevLett,
  title = {Photonic Topological Insulating Phase Induced Solely by Gain and Loss},
  author = {Takata, Kenta and Notomi, Masaya},
  journal = {Phys. Rev. Lett.},
  volume = {121},
  issue = {21},
  pages = {213902},
  numpages = {6},
  year = {2018},
  month = {Nov},
  publisher = {American Physical Society},
  doi = {10.1103/PhysRevLett.121.213902},
  url = {https://link.aps.org/doi/10.1103/PhysRevLett.121.213902}
}

@article{Teo2022PhyRevA,
  title = {Topological phase transition induced by gain and loss in a photonic Chern insulator},
  author = {Teo, Hau Tian and Xue, Haoran and Zhang, Baile},
  journal = {Phys. Rev. A},
  volume = {105},
  issue = {5},
  pages = {053510},
  numpages = {5},
  year = {2022},
  month = {May},
  publisher = {American Physical Society},
  doi = {10.1103/PhysRevA.105.053510},
  url = {https://link.aps.org/doi/10.1103/PhysRevA.105.053510}
}

@article{Liu2020PhysRevAppl,
  title = {Gain- and Loss-Induced Topological Insulating Phase in a Non-Hermitian Electrical Circuit},
  author = {Liu, Shuo and Ma, Shaojie and Yang, Cheng and Zhang, Lei and Gao, Wenlong and Xiang, Yuan Jiang and Cui, Tie Jun and Zhang, Shuang},
  journal = {Phys. Rev. Appl.},
  volume = {13},
  issue = {1},
  pages = {014047},
  numpages = {11},
  year = {2020},
  month = {Jan},
  publisher = {American Physical Society},
  doi = {10.1103/PhysRevApplied.13.014047},
  url = {https://link.aps.org/doi/10.1103/PhysRevApplied.13.014047}
}

@article{Nasari2023OptMaterExpress,
author = {Hadiseh Nasari and Georgios G. Pyrialakos and Demetrios N. Christodoulides and Mercedeh Khajavikhan},
journal = {Opt. Mater. Express},
number = {4},
pages = {870--885},
publisher = {Optica Publishing Group},
title = {Non-Hermitian topological photonics},
volume = {13},
month = {Apr},
year = {2023},
url = {https://opg.optica.org/ome/abstract.cfm?URI=ome-13-4-870},
doi = {10.1364/OME.483361}
}

@article{Hashemi2025APLPhotonics,
    author = {Hashemi, Amin and Zakeri, M. Javad and Jung, Pawel S. and Blanco-Redondo, Andrea},
    title = {Topological quantum photonics},
    journal = {APL Photonics},
    volume = {10},
    number = {1},
    pages = {010903},
    year = {2025},
    month = {01},
    issn = {2378-0967},
    doi = {10.1063/5.0239265},
    url = {https://doi.org/10.1063/5.0239265}
}

@article{Price2022JPPhotonics,
doi = {10.1088/2515-7647/ac4ee4},
url = {https://dx.doi.org/10.1088/2515-7647/ac4ee4},
year = {2022},
month = {jun},
publisher = {IOP Publishing},
volume = {4},
number = {3},
pages = {032501},
author = {Price, Hannah and Chong, Yidong and Khanikaev, Alexander and Schomerus, Henning and Maczewsky, Lukas J and Kremer, Mark and Heinrich, Matthias and Szameit, Alexander and Zilberberg, Oded and Yang, Yihao and Zhang, Baile and Alù, Andrea and Thomale, Ronny and Carusotto, Iacopo and St-Jean, Philippe and Amo, Alberto and Dutt, Avik and Yuan, Luqi and Fan, Shanhui and Yin, Xuefan and Peng, Chao and Ozawa, Tomoki and Blanco-Redondo, Andrea},
title = {Roadmap on topological photonics},
journal = {Journal of Physics: Photonics}
}

@article{Yan2023Nanophotonics,
url = {https://doi.org/10.1515/nanoph-2022-0775},
title = {Advances and applications on non-Hermitian topological photonics},
title = {},
author = {Qiuchen Yan and Boheng Zhao and Rong Zhou and Rui Ma and Qinghong Lyu and Saisai Chu and Xiaoyong Hu and Qihuang Gong},
pages = {2247--2271},
volume = {12},
number = {13},
journal = {Nanophotonics},
doi = {doi:10.1515/nanoph-2022-0775},
year = {2023},
lastchecked = {2025-03-25}
}

@article{Hatano1996PhysRevLett,
  title = {Localization Transitions in Non-Hermitian Quantum Mechanics},
  author = {Hatano, Naomichi and Nelson, David R.},
  journal = {Phys. Rev. Lett.},
  volume = {77},
  issue = {3},
  pages = {570--573},
  numpages = {0},
  year = {1996},
  month = {Jul},
  publisher = {American Physical Society},
  doi = {10.1103/PhysRevLett.77.570},
  url = {https://link.aps.org/doi/10.1103/PhysRevLett.77.570}
}

@article{Yao2018PhysRevLett,
  title = {Edge States and Topological Invariants of Non-Hermitian Systems},
  author = {Yao, Shunyu and Wang, Zhong},
  journal = {Phys. Rev. Lett.},
  volume = {121},
  issue = {8},
  pages = {086803},
  numpages = {8},
  year = {2018},
  month = {Aug},
  publisher = {American Physical Society},
  doi = {10.1103/PhysRevLett.121.086803},
  url = {https://link.aps.org/doi/10.1103/PhysRevLett.121.086803}
}

@article{Zhang31122022,
author = {Xiujuan Zhang and Tian Zhang and Ming-Hui Lu and Yan-Feng Chen and},
title = {A review on non-Hermitian skin effect},
journal = {Advances in Physics: X},
volume = {7},
number = {1},
pages = {2109431},
year = {2022},
publisher = {Taylor \& Francis},
doi = {10.1080/23746149.2022.2109431},
URL = { https://doi.org/10.1080/23746149.2022.2109431}
}

@article{Andrea2016PhysRevLett,
  title = {Topological Optical Waveguiding in Silicon and the Transition between Topological and Trivial Defect States},
  author = {Blanco-Redondo, Andrea and Andonegui, Imanol and Collins, Matthew J. and Harari, Gal and Lumer, Yaakov and Rechtsman, Mikael C. and Eggleton, Benjamin J. and Segev, Mordechai},
  journal = {Phys. Rev. Lett.},
  volume = {116},
  issue = {16},
  pages = {163901},
  numpages = {5},
  year = {2016},
  month = {Apr},
  publisher = {American Physical Society},
  doi = {10.1103/PhysRevLett.116.163901},
  url = {https://link.aps.org/doi/10.1103/PhysRevLett.116.163901}
}

@article{
doi:10.1126/science.aau4296,
author = {Andrea Blanco-Redondo  and Bryn Bell  and Dikla Oren  and Benjamin J. Eggleton  and Mordechai Segev },
title = {Topological protection of biphoton states},
journal = {Science},
volume = {362},
number = {6414},
pages = {568-571},
year = {2018},
doi = {10.1126/science.aau4296},
URL = {https://www.science.org/doi/abs/10.1126/science.aau4296}
}

@article{PRXQuantum.6.010338,
  title = {Versatile Chip-Scale Platform for High-Rate Entanglement Generation Using an $\mathrm{Al}\mathrm{Ga}\mathrm{As}$ Microresonator Array},
  author = {Pang, Yiming and Castro, Joshua E. and Steiner, Trevor J. and Duan, Liao and Tagliavacche, Noemi and Borghi, Massimo and Thiel, Lillian and Lewis, Nicholas and Bowers, John E. and Liscidini, Marco and Moody, Galan},
  journal = {PRX Quantum},
  volume = {6},
  issue = {1},
  pages = {010338},
  numpages = {15},
  year = {2025},
  month = {Mar},
  publisher = {American Physical Society},
  doi = {10.1103/PRXQuantum.6.010338},
  url = {https://link.aps.org/doi/10.1103/PRXQuantum.6.010338}
}

@article{
doi:10.1126/science.aar7709,
author = {Mohammad-Ali Miri  and Andrea Alù },
title = {Exceptional points in optics and photonics},
journal = {Science},
volume = {363},
number = {6422},
pages = {eaar7709},
year = {2019},
doi = {10.1126/science.aar7709},
URL = {https://www.science.org/doi/abs/10.1126/science.aar7709}
}

@book{Vahalabook,
author = {Vahala, Kerry},
title = {Optical Microcavities},
publisher = {WORLD SCIENTIFIC},
year = {2004},
doi = {10.1142/5485},
address = {},
edition   = {},
URL = {https://www.worldscientific.com/doi/abs/10.1142/5485}
}

@book{Vanbook,
author = {Vien Van},
title = {Optical Microring Resonators},
publisher = {CRC Press},
year = {2017},
doi = {https://doi.org/10.1201/9781315303512}
}

@article{Baranov2017Optica,
author = {Denis G. Baranov and Alex Krasnok and Andrea Al\`{u}},
journal = {Optica},
number = {12},
pages = {1457--1461},
publisher = {Optica Publishing Group},
title = {Coherent virtual absorption based on complex zero excitation for ideal light capturing},
volume = {4},
month = {Dec},
year = {2017},
url = {https://opg.optica.org/optica/abstract.cfm?URI=optica-4-12-1457},
doi = {10.1364/OPTICA.4.001457}
}

@article{Longhi2018OptLett,
author = {S. Longhi},
journal = {Opt. Lett.},
number = {9},
pages = {2122--2125},
publisher = {Optica Publishing Group},
title = {Coherent virtual absorption for discretized light},
volume = {43},
month = {May},
year = {2018},
url = {https://opg.optica.org/ol/abstract.cfm?URI=ol-43-9-2122},
doi = {10.1364/OL.43.002122}
}

@article{Zhong2020PhysRevResearch,
  title = {Coherent virtual absorption of light in microring resonators},
  author = {Zhong, Q. and Simonson, L. and Kottos, T. and El-Ganainy, R.},
  journal = {Phys. Rev. Res.},
  volume = {2},
  issue = {1},
  pages = {013362},
  numpages = {6},
  year = {2020},
  month = {Mar},
  publisher = {American Physical Society},
  doi = {10.1103/PhysRevResearch.2.013362},
  url = {https://link.aps.org/doi/10.1103/PhysRevResearch.2.013362}
}

@book{stochasticprocessesbook,
author = {Paul, Wolfgang and Baschnagel, Jorg},
title = {Stochastic Processes: From Physics to Finance},
publisher = {Springer},
year = {2013},
doi = {},
address = {},
edition   = {},
URL = {https://link.springer.com/book/10.1007/978-3-319-00327-6}
}

@book{SDEbook,
author = {Bernt {\O}ksendal},
title = {Stochastic Differential Equations},
publisher = {Springer},
year = {2003},
doi = {},
address = {},
edition   = {},
URL = {https://link.springer.com/book/10.1007/978-3-642-14394-6}
}

@article{Jorge2024Advanced_photonics_nexus,
	author = {Jorge Parra and Juan Navarro-Arenas and Pablo Sanchis},
	title = {Silicon thermo-optic phase shifters: a review of configurations and optimization strategies},
	journal = {Advanced Photonics Nexus},
	year = {2024},
	volume = {3},
	number = {4},
	month = {May},
	doi = {10.1117/1.apn.3.4.044001},
	date = {2024-05-24},
	url = {https://lens.org/051-386-829-011-364}
}

@misc{hashemi2026programmableonchipsynthesisreconstruction,
      title={Programmable on-chip synthesis and reconstruction of partially coherent two-mode optical fields}, 
      author={Amin Hashemi and Abbas Shiri and Bahaa E. A. Saleh and Andrea Blanco-Redondo and Ayman F. Abouraddy},
      year={2026},
      eprint={2601.09802},
      archivePrefix={arXiv},
      primaryClass={physics.optics},
      url={https://arxiv.org/abs/2601.09802}, 
}

@misc{hashemi2026onchipcontrolcoherencematrix,
      title={On-chip control of the coherence matrix of four-mode partially coherent light: rank, entropy, and modal Stokes parameters}, 
      author={Amin Hashemi and Abbas Shiri and Bahaa E. A. Saleh and Andrea Blanco-Redondo and Ayman F. Abouraddy},
      year={2026},
      eprint={2601.18797},
      archivePrefix={arXiv},
      primaryClass={physics.optics},
      url={https://arxiv.org/abs/2601.18797}, 
}

@book{statisticalopticsbook,
author = {Joseph W. Goodman},
title = {Statistical Optics},
publisher = {Wiley},
year = {2015},
doi = {},
address = {},
edition   = {second},
}

@article{RevModPhys.37.231,
  title = {Coherence Properties of Optical Fields},
  author = {MANDEL, L. and WOLF, E.},
  journal = {Rev. Mod. Phys.},
  volume = {37},
  issue = {2},
  pages = {231--287},
  numpages = {0},
  year = {1965},
  month = {Apr},
  publisher = {American Physical Society},
  doi = {10.1103/RevModPhys.37.231},
  url = {https://link.aps.org/doi/10.1103/RevModPhys.37.231}
}

@article{Vinzenz2025PRA,
  title = {Coherence-independent non-Hermitian topological filters},
  author = {Zimmermann, Vinzenz and Hashemi, Amin and Busch, Kurt and Blanco-Redondo, Andrea and Perez-Leija, Armando},
  journal = {Phys. Rev. A},
  volume = {112},
  issue = {6},
  pages = {063519},
  numpages = {7},
  year = {2025},
  month = {Dec},
  publisher = {American Physical Society},
  doi = {10.1103/ylwg-dh7t},
  url = {https://link.aps.org/doi/10.1103/ylwg-dh7t}
}

@book{MonteCarloSimulationBool,
author = {Binder, Kurt and Heermann, Dieter W. },
title = {Monte Carlo Simulation in Statistical Physics},
publisher = {Springer},
year = {2010},
doi = {},
address = {},
edition   = {5th},
}

\end{document}